\documentclass[prl, twocolumn,amsmath,amssymb,superscriptaddress]{revtex4}
\usepackage{graphicx, subfigure}
\usepackage{dcolumn}
\usepackage{color}
\usepackage{bm}
\usepackage[T1]{fontenc}
\usepackage[matrix,frame,arrow]{xypic}

\newcommand{\bket}[1]{\left<#1\right>}
\newcommand{\abs}[1]{\left|#1\right|}

\newcommand{\ket}[1]{\left|#1\right>}
\newcommand{\bra}[1]{\left<#1\right|}
\newcommand{\tr}[1]{\text{Tr}\left(#1\right)}

\newcommand{\ba}{\text{\bf{a}}}

\newcommand{\bD}{\text{\bf{D}}}
\newcommand{\bP}{\text{\bf{P}}}
\newcommand{\bH}{\text{\bf{H}}}

\newcommand{\bphi}{\text{\boldsymbol{$\varphi$}}}
\newcommand{\norm}[1]{\left||#1\right||}
\newcommand{\bA}{\text{\bf{A}}}

\newcommand{\NN}{{\mathcal N}}

\newcommand{\beginsupplement}{%
        \setcounter{table}{0}
        \renewcommand{\thetable}{S\arabic{table}}%
        \setcounter{figure}{0}
        \renewcommand{\thefigure}{S\arabic{figure}}%
        \setcounter{equation}{0}
        \renewcommand{\theequation}{S\arabic{equation}}
     }

\begin{document}

\title{Confining the state of light to a quantum manifold by engineered two-photon loss}
\author{Zaki Leghtas}
\affiliation{Department of Applied Physics, Yale University, New Haven, Connecticut 06520, USA}
\author{Steven Touzard}
\affiliation{Department of Applied Physics, Yale University, New Haven, Connecticut 06520, USA}
\author{Ioan M. Pop}
\affiliation{Department of Applied Physics, Yale University, New Haven, Connecticut 06520, USA}
\author{Angela Kou}
\affiliation{Department of Applied Physics, Yale University, New Haven, Connecticut 06520, USA}
\author{Brian Vlastakis}
\affiliation{Department of Applied Physics, Yale University, New Haven, Connecticut 06520, USA}
\author{Andrei Petrenko}
\affiliation{Department of Applied Physics, Yale University, New Haven, Connecticut 06520, USA}
\author{Katrina M. Sliwa}
\affiliation{Department of Applied Physics, Yale University, New Haven, Connecticut 06520, USA}
\author{Anirudh Narla}
\affiliation{Department of Applied Physics, Yale University, New Haven, Connecticut 06520, USA}
\author{Shyam Shankar}
\affiliation{Department of Applied Physics, Yale University, New Haven, Connecticut 06520, USA}
\author{Michael J. Hatridge}
\affiliation{Department of Applied Physics, Yale University, New Haven, Connecticut 06520, USA}
\author{Matthew Reagor}
\affiliation{Department of Applied Physics, Yale University, New Haven, Connecticut 06520, USA}
\author{Luigi Frunzio}
\affiliation{Department of Applied Physics, Yale University, New Haven, Connecticut 06520, USA}
\author{Robert J. Schoelkopf}
\affiliation{Department of Applied Physics, Yale University, New Haven, Connecticut 06520, USA}
\author{Mazyar Mirrahimi}
\affiliation{INRIA Paris-Rocquencourt, Domaine de Voluceau, B.P.~105, 78153 Le Chesnay Cedex, France}
\affiliation{Department of Applied Physics, Yale University, New Haven, Connecticut 06520, USA}
\author{Michel H. Devoret}
\affiliation{Department of Applied Physics, Yale University, New Haven, Connecticut 06520, USA}

\date{\today}

\begin{abstract}
Physical systems usually exhibit quantum behavior, such as superpositions and entanglement, only when they are sufficiently decoupled from a lossy environment. Paradoxically, a specially engineered interaction with the environment can become a resource for the generation and protection of quantum states. This notion can be generalized to the confinement of a system into a manifold of quantum states, consisting of all coherent superpositions of multiple stable steady states. We have experimentally confined the state of a harmonic oscillator to the quantum manifold spanned by two coherent states of opposite phases. In particular, we have observed a Schr\"{o}dinger cat state spontaneously squeeze out of vacuum, before decaying into a classical mixture. This was accomplished by designing a superconducting microwave resonator whose coupling to a cold bath is dominated by photon pair exchange. This experiment opens new avenues in the fields of nonlinear quantum optics and quantum information, where systems with multi-dimensional steady state manifolds can be used as error corrected logical qubits.
\end{abstract}
\maketitle

{Maintaining the state of a system in the vicinity of a predefined state despite the presence of external perturbations plays a central role in science and engineering. Examples of this notion, called stabilization, include devices such as Watt's governor for regulating the angular velocity in a steam engine, and the escapement mechanism which prevents decay of the oscillation of a pendulum in clocks. The problem of stabilizing a quantum system is fundamentally more subtle than stabilizing a classical one. Stabilizing a system requires an interaction which, quantum mechanically, is always invasive. The mere act of learning something about a system perturbs it. Carefully designed non destructive quantum measurements have recently been incorporated in feedback loops to stabilize a \emph{single} quantum state \cite{Sayrin2011,Vijay2012,Riste2012,Campagne-Ibarcq2013}. Alternatively, adequately engineering an interaction with an auxiliary dissipative system, termed engineered dissipation,  can also stabilize a single quantum state \cite{Krauter2011,Murch2012,Geerlings2013,Shankar2013,Lin2013}. }

{Can engineered dissipation protect \emph{all} unknown superpositions of two states, thus protecting quantum information? In fact, the static random access memory of a computer chip dynamically stabilizes the states representing 0 and 1 by combining the energy supply and dissipation, providing fast access time and robustness against noise. For a quantum memory, however, one must construct a system with not only one or two stable steady states (SSSs), but rather a whole quantum manifold composed of {all} coherent superpositions of two SSSs, see Fig.~\ref{fig:schema}a. By construction, such a system does not distinguish between all its SSSs and hence cannot correct for errors within the SSS manifold. However, quantum information encoded in this manifold will be protected against perturbations which move it out of the manifold.}

{An oscillator which exchanges only pairs of photons with a dissipative auxiliary system \cite{Wolinsky1988} is a practical example which displays a manifold of SSSs. This two-photon loss will force and confine the state of the oscillator into the quantum manifold spanned by two oscillation states with opposite phases. Uncontrolled energy decay, termed single photon loss, causes decoherence within the SSS manifold, and hence quantum superpositions will eventually decay into classical mixtures. Nevertheless, in the regime where pairs of photons are extracted at a rate at least as large as the single photon decay rate, transient quantum coherence can be observed.}

{This regime, essential to the proposal by Wolinsky and Carmichael \cite{Wolinsky1988},  had not been reached as it requires combining strong non-linear interactions between modes and low single photon decay rates. Our experiment enters this regime through a circuit quantum electrodynamics (cQED) architecture \cite{Wallraff2004}, benefiting from the strong non-linearity and low loss of a Josephson junction.} Our setup, schematically described in Fig.~\ref{fig:schema}b, is based on a recent proposal \cite{Mirrahimi2014}. It consists of two superconducting microwave oscillators coupled through a Josephson junction in a bridge transmon configuration \cite{Kirchmair2013}. These oscillators are the fundamental modes of two superconducting cavities. One cavity, termed the storage, holds the manifold of SSSs and is designed to have minimal {single photon} dissipation. The other, termed the readout, is over-coupled to a transmission line and its role is to evacuate entropy from the storage. {In a variety of non-linear systems, {the} interaction of a pump tone with relevant degrees of freedom provides cooling \cite{Teufel2011}, squeezing \cite{Drummond2010}, and amplification \cite{Siddiqi2004,Castellanos-Beltran2008}}. Similarly, we use the four-wave mixing capability of the Josephson junction {to} generate a {coupling which exchanges pairs of photon in the storage with single photons in the readout}. 

By off-resonantly pumping the readout at an angular frequency
\begin{equation}
\omega_p=2\omega_s-\omega_r\;,
\label{eq:freqmatching}
\end{equation}
where $\omega_{r,s}$ are the readout and storage {angular} frequencies respectively, {the pump stimulates the conversion of two storage photons into one readout and one pump photon. The readout photon then rapidly dissipates through the transmission line, resulting in a loss in photon-pairs for the storage, as illustrated in Fig.~\ref{fig:schema}d}. This engineered dissipation is the key ingredient in our experiment. The input power that balances this dissipation is provided by {the readout drive}: a weak resonant irradiation of the readout. Due to {the non-linear mixing with the pump, these input readout photons are converted into pairs of storage photons, as illustrated in Fig.~\ref{fig:schema}e}. {Unlike the usual linear driven-dissipative oscillator which adopts only one oscillation state, our non-linear driven-dissipative system displays a quantum manifold of SSSs corresponding to all superpositions of two oscillation states with opposite phases.} 


In the experiment, we employed a third mode besides the storage and readout: the excitation of the bridge transmon qubit, restricted to its ground and first excited state. It served as a calibration tool for all the experimental parameters, and as a means to directly measure the Wigner function of the storage.

{Our system is well described by the effective Hamiltonian for the storage and the readout \cite{supp}:}
\begin{eqnarray}
\bH_{sr}/\hbar&=&g_2^*\ba_s^2\ba_r^\dag+g_2(\ba_s^\dag)^2\ba_r+\epsilon_d\ba_r^\dag+\epsilon_d^*\ba_r\notag\\
&-&\chi_{rs}\ba_r^\dag\ba_r\ba_s^\dag\ba_s-\sum_{m=r,s}\frac{\chi_{mm}}{2}{\ba_m^\dag}^2\ba_m^2\;.
\label{eq:twomodesH}
\end{eqnarray}
The readout and storage annihilation operators are denoted $\ba_{r}$ and $\ba_s$ respectively. The first line is a microscopic Hamiltonian of the degenerate parametric oscillator \cite{Carmichael2007} with
$$
g_2=\chi_{rs}\xi_p^*/2\;,\qquad \xi_p\approx{-i\epsilon_p}/\left({\frac{\kappa_r}{2}+i(\omega_r-\omega_p)}\right)\;,
$$
where $\chi_{rs}/2\pi=206~$kHz is the dispersive coupling between the readout and the storage, and $\epsilon_{p}, \epsilon_d$ are the pump and drive amplitudes, respectively. The terms in $g_2$ correspond to the conversion of pairs of photons in the storage into single photons in the readout (Fig.~\ref{fig:schema}d-e). The readout and storage have a Kerr non-linearity: $\chi_{rr}/2\pi=2.14~$MHz and $\chi_{ss}/2\pi\approx4~$kHz, respectively. The Kerr interactions can be considered as perturbations which do not significantly disturb the two-photon conversion effects \cite{supp}. The storage and readout single photon lifetimes are respectively $1/\kappa_s=20~\mu$s and $1/\kappa_r=25~$ns.


The two-photon processes shown in Fig.~\ref{fig:schema}d-e are only activated when the frequency matching condition \eqref{eq:freqmatching} is met. We satisfy this condition by performing a calibration experiment {as shown} in Fig.~\ref{fig:readoutspec_vs_pumpfreq}. We excited the readout with a weak CW probe tone ($\approx1$ photon), and measured its transmitted power, in presence of the pump tone, while sweeping the frequency of both tones. The pump power is kept fixed during this measurement, and its value was chosen as the largest that did not degrade the coherence times of our system  \cite{supp}. When the frequency matching condition is met, the probe photons are converted back and forth into pairs of storage photons (Fig.~\ref{fig:schema}d-e). When equilibrium is reached for this process, the input probe photons interfere destructively with the back-converted storage photons and are now reflected {back into the probe input port} \cite[Section 12.1.1]{Carmichael2007}: the readout is in an induced dark state. The {dip} in Fig.~\ref{fig:readoutspec_vs_pumpfreq}(a-b) is a signature of this {interference}. {Its depth indicates that we have achieved a large non-linear coupling $g_2\gg\kappa_s$  \cite{supp}}. For the subsequent experiments, we fixed the pump frequency to $\omega_p/2\pi=8.011~$GHz, {which makes the dip coincide with the readout resonance frequency}. 

We demonstrate that photons are inserted in the storage by measuring the {probability of having $n>0$ photons in the storage while sweeping} the readout drive frequency, as shown in Fig.~\ref{fig:readoutspec_vs_pumpfreq}(c). {We apply a 10 $\mu$s square pulse from the pump and drive tones simultaneously, and then excite the qubit from its ground to its excited state, conditioned on there being $n=0$ photons in the storage \cite{Johnson2010}. Reading out the qubit state then answers the question: are there 0 photons in the storage?} The peak at zero detuning shows that the readout drive and the pump combine non-linearly to insert photons {into} the storage. {We then choose the drive tone frequency which maximizes the number of photons in the storage,} and the drive power is fixed to ensure an equilibrium average photon number in the storage of $\approx 4$  \cite{supp}.

Adiabatically eliminating the readout from \eqref{eq:twomodesH} \cite{supp,Carmichael2007}, we obtain a dynamics for the storage governed by the Hamiltonian
\begin{equation*}
\bH_s/\hbar=\epsilon_2^*\ba_s^2+\epsilon_2(\ba_s^\dag)^2-\frac{\chi_{ss}}{2}{\ba_s^\dag}^2\ba_s^2\;,
\end{equation*}
and loss operators $\sqrt{\kappa_2}\ba_s^2$ and $\sqrt{\kappa_s}\ba_s$, where
$$
\epsilon_2=-i\frac{\chi_{sr}}{\kappa_r}\xi_p^*\epsilon_d\;,\qquad \kappa_2=\frac{\chi_{sr}^2}{\kappa_r}\abs{\xi_p}^2\;.
$$
The $\epsilon_2$  non-linear drive inserts pairs of photons in the storage (Fig.~\ref{fig:schema}e) and is analogous to the usual squeezing drive of a non-linear oscillator \cite{Drummond2010}. The novel element in this experiment is the non-linear decay, of rate $\kappa_2$, which extracts only photons in pairs from the storage (Fig.~\ref{fig:schema}d). In absence of unavoidable loss $\kappa_s$ and neglecting the effect of $\chi_{ss}$ \cite{supp}, the storage converges into the two-dimensional quantum manifold spanned by coherent states $\ket{\pm\alpha_\infty}$, where
\begin{equation*}
\alpha_\infty\Big|_{\chi_{ss}=\kappa_s=0}=i\sqrt{\frac{2\epsilon_d}{\xi_p\chi_{sr}}}\;.
\end{equation*}

In a classical model where quantum noise is just ordinary noise \cite{DYKMAN1980}, our system behaves as a bi-stable oscillator with two oscillation states of amplitudes $\pm\alpha_\infty$. The storage then evolves to $+\alpha_\infty$ \emph{or} $-\alpha_\infty$. However in the full quantum model, the storage must evolve to $+\alpha_\infty$ \emph{and} $-\alpha_\infty$ when initialized in the vacuum state, thus forming an even Schr\"{o}dinger cat state: $\NN(\ket{\alpha_\infty}+\ket{-\alpha_\infty})=\NN(\sum_{n=0}^\infty{(\alpha_\infty^{2n}/{2n!})\ket{2n}})$  ($\NN$ is a normalization constant) \cite{Ourjoumtsev2007,Vlastakis2013,Deleglise2008,Monroe1996,Hofheinz2009}.

 {We visualize these dynamics by measuring the state of the storage by direct Wigner tomography \cite{Lutterbach1997,Vlastakis2013}. The Wigner function \cite{Haroche2006} is a representation of a quantum state defined over the complex plane as $W(\alpha)=\frac{2}{\pi}\bket{\bD_\alpha \bP \bD_{-\alpha}}$, the normalized expectation value of the parity operator $\bP=e^{i\pi\ba_s^\dag\ba_s}$ for the state displaced by the operator $\bD_\alpha=e^{\alpha \ba_s^\dag-\alpha^* \ba_s}$.} This quasi-probability distribution vividly displays the quantum features of a coherent superposition.

The bi-stable property of our system is demonstrated in Fig.~\ref{fig:double_well} by initializing the storage in coherent states with a mean photon number of 6.8 with various phases, and observing their convergence to the closest equilibrium state (Fig.~\ref{fig:double_well}, {displacement angle = \{0,$\pm \pi/4$,$\pm 3\pi/4$,$\pi$\}}). The upper and lower middle panels (Fig.~\ref{fig:double_well}~, {displacement angle = $\pm\pi/2$}) correspond to states initialized at almost equal distance from $\pm\alpha_\infty$ which {randomly evolve} to one equilibrium state or the other, thus converging to the statistical mixture of $\pm\alpha_\infty$.


The coherent splitting of the vacuum into the quantum superposition of $\ket{\pm\alpha_\infty}$ is demonstrated in Fig.~\ref{fig:time_evolution}. In absence of loss in the storage, the pairwise exchange of photons between the storage and the environment conserves parity. Therefore, since the vacuum state is an even parity state, it must transform into the even cat state: the unique even state contained in the manifold of equilibrium states. Similarly, Fock state $\ket{1}$ being an odd parity state, it must transform into the odd cat state  \cite{supp}. In presence of $\kappa_s$, all coherences will ultimately disappear. However, for large enough $\kappa_2$, a {quantum superposition} transient state is observed. 


In this experiment, we achieve $\abs{\xi_p}^2=1.2$ which implies {$g_2/2\pi=111~$kHz} and {$\kappa_2/\kappa_s=1.0$}. The quantum {nature} of the transient storage state is visible in the negative fringes of the Wigner function (see Fig.~\ref{fig:time_evolution}a-b,$7~\mu$s), and the non-Poissonian photon number statistics (Fig.~\ref{fig:time_evolution}d,$7~\mu$s). {After 7 $\mu$s of pumping, we obtain a state with an average photon number $\bar n=2.4$, and a parity of 42$\%$, which is larger than the parity of a thermal state ($17\%$) or a coherent state (0.8$\%$) with equal $\bar n$. After 19$\mu$s of pumping, although the negative fringes vanish, the phase and amplitude of the SSSs $\ket{\pm\alpha_\infty}$ are conserved. Our data is in good agreement with numerical simulations, see Fig.~\ref{fig:time_evolution}c, {indicating that our dominant source of imperfection is single photon loss}. {These results illustrate the confinement of the storage state into the manifold of SSSs, and how it transits through a quantum superposition of $\ket{\pm\alpha_\infty}$.}

In conclusion, we have realized a non-linearly driven-dissipative oscillator which spontaneously evolves towards the quantum manifold spanned by two coherent states. Starting from the vacuum, a Schr\"{o}dinger cat state is produced, as shown by negativities in the Wigner function and a non-Poissonian photon number distribution. This was achieved by attaining the regime in which the photon pair exchange rate is of the same order as the single photon decay rate. The ratio between these two rates can be further improved within the present technology by using a higher $Q$ oscillator and increasing its non-linear coupling to the bath.
Our experiment is an essential step towards a new paradigm for universal quantum computation \cite{Mirrahimi2014}. By combining higher order forms of our non-linear dissipation with efficient error syndrome measurements \cite{Sun2014}, quantum information can be encoded and manipulated in a protected manifold of quantum states.


{\textbf{Acknowledgements}: The authors thank L. Jiang and V. V. Albert for helpful discussions. Facilities use was supported by YINQE and NSF MRSEC DMR 1119826. This research was supported by  ARO under Grant No. W911NF-14-1-0011. MM acknowledges support from the French ``Agence Nationale de la Recherche'' under the project EPOQ2 number ANR-09-JCJC-0070.}

\begin{figure*}[ht!]
\setlength{\unitlength}{1cm}
{\includegraphics[width=1.5\columnwidth]{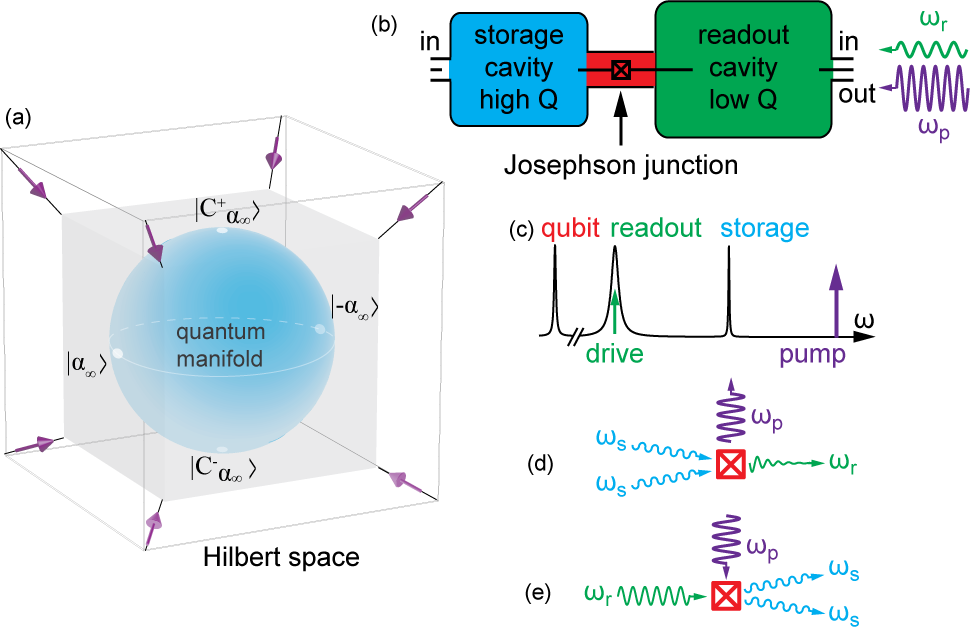}}
\caption{Schematic of the experiment. {(a) Confinement of a quantum state belonging to a large Hilbert space into a two-dimensional quantum manifold. The outer and inner cubes form a hypercube representing a multi-dimensional Hilbert space. The inner blue sphere represents the manifold of states spanned by the two coherent states $\ket{\pm\alpha_\infty}$. Quantum states such as the even and odd Schr\"{o}dinger cat states $\ket{C^\pm_{\alpha_\infty}}=\NN\left({\ket{\alpha_\infty}\pm\ket{-\alpha_\infty}}\right)$ also belong to this manifold, where $\NN$ is a normalization factor. Stabilizing forces direct all states towards the inner sphere without inducing any rotation in this subspace, as indicated by the purple arrows.} (b) Two superconducting cavities are coupled through a Josephson junction. Pump and drive microwave tones are applied to the readout, creating the appropriate nonlinear interaction which generates a coherent superposition of steady states in the storage. The readout output port is connected to an amplifier chain \cite{supp}. Direct Wigner tomography of the storage is performed using its input port and the qubit mode. (c) Schematic representation of the spectrum of different modes involved in the experiment. The pump and drive tones are shown as vertical arrows. (d-e) Four-wave processes involved in the nonlinear damping and nonlinear drive, respectively, experienced by the storage. In (d), two photons of the storage combine and convert, stimulated by the pump tone, into a readout photon which is irreversibly radiated away by the transmission line. This process is balanced by the conversion of the drive tone, which in presence of the pump, creates two photons in the storage (e).}
\label{fig:schema}
\end{figure*}
%
\begin{figure*}
\setlength{\unitlength}{1cm}
\includegraphics[width=1.5\columnwidth]{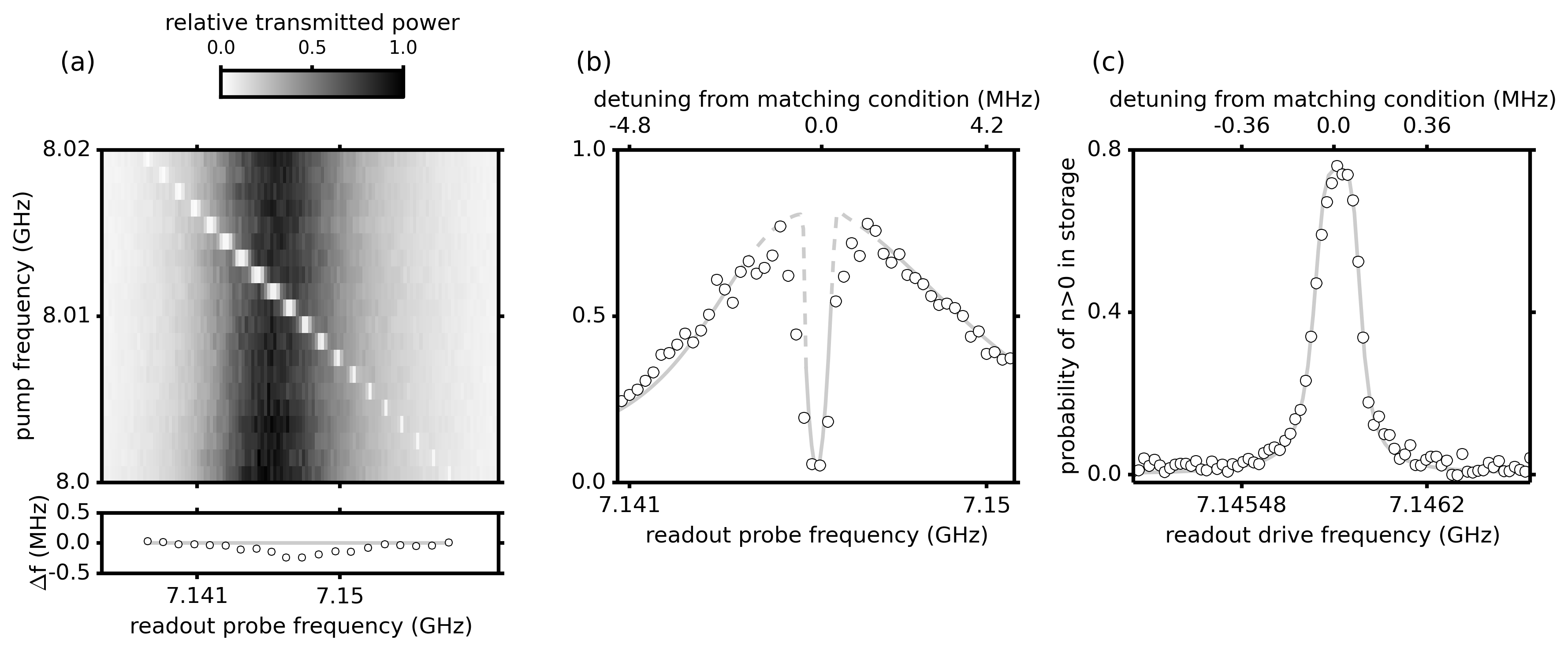}
\caption{(a-b) CW spectroscopy of the readout in presence of the pump tone. The grey-scale represents transmitted power of the probe tone through the readout as a function of probe frequency (horizontal axis) and pump frequency (vertical axis). In the {top panel of} (a), the usual Lorentzian response develops a sharp and deep dip signaling conversion of probe photons into storage photons. The dip frequency $\omega_\text{dip}(\omega_p)$ decreases as the pump frequency increases. In the lower panel of (a), we plot for each dip, $\Delta f=\omega_\text{dip}/2\pi-(2\omega_s-\omega_p)/2\pi$, and we see that the deviation of the data (open dots) to the theory (full line: $\Delta f=0$) is only of the order of 0.24 MHz over a span of 20 MHz. Note that here, we use the Stark-shifted value of $\omega_s$ due to the pump  \cite{supp}. (b) Cut of the grey-scale map (a) at $\omega_p/2\pi=8.011~$GHz. (c) Conversion seen from the storage, represented as probability of not being in the vacuum state, as a function of drive frequency. The dashed and full lines in (b-c) are the result of a numerical computation of the steady state density matrix of the system with Hamiltonian \eqref{eq:twomodesH} and loss operators $\sqrt{\kappa_s}\ba_s, \sqrt{\kappa_r}\ba_r$, sweeping the drive frequency and keeping the pump frequency fixed. All parameters entering in theoretical predictions were measured or estimated independently. However in (c), the theory was rescaled by a factor of 0.76 to fit the data. We believe that the need for this rescaling is a consequence of the unexplained modified qubit relaxation times {when} the pump and the drive are on  \cite{supp}.}
\label{fig:readoutspec_vs_pumpfreq}
\end{figure*}

\begin{figure*}
\setlength{\unitlength}{1cm}
{\includegraphics[width=1.2\columnwidth]{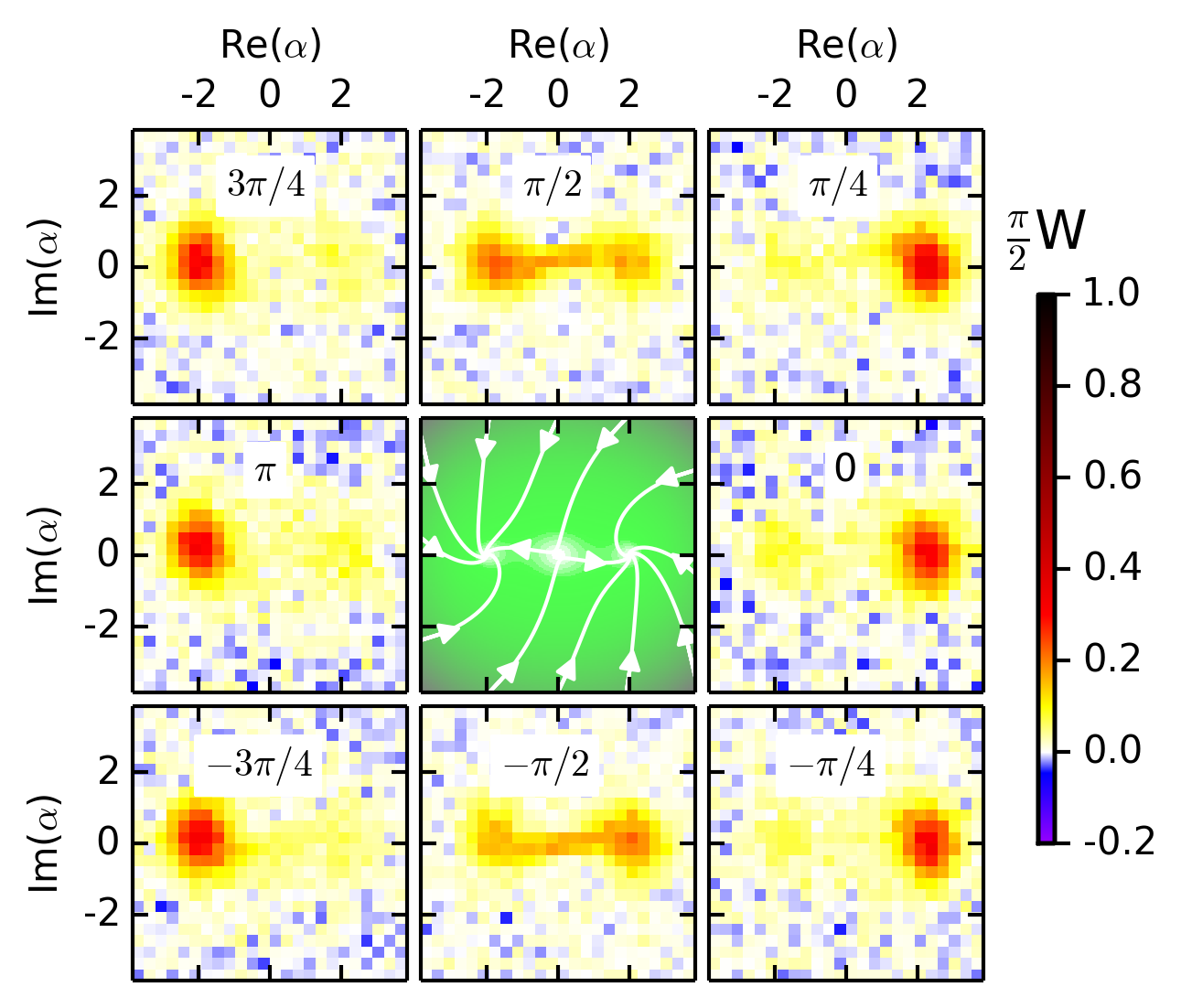}}
\caption{Bi-stable behavior of the steady state manifold of the non-linearly driven-damped storage oscillator. The central panel shows the theoretical classical equivalent of a potential of the storage nonlinear dynamics. The modulus of the velocity (color) has three zeros corresponding to two SSSs $\ket{\pm\alpha_\infty}$ and the saddle point $\ket{0}$. Trajectories initialized on the panel border converge to one of these two SSSs. {These trajectories are curved due to the Kerr effect}. The outside panels show the measured Wigner function of the storage after $10~\mu$s of pumping for different initial states. For each panel, we initialize the storage in a coherent state of amplitude $\alpha_k$, where $\abs{\alpha_k}=2.6$ and $\arg(\alpha_k)$ is indicated in each panel. The storage converges to a combination of $\ket{\pm\alpha_\infty}$. The weight of each of these two states and the coherence of their superposition is set by the initial state. For the initial phases $\arg(\alpha_k)=0,\pm\pi/4$, the storage mainly evolves to $\ket{\alpha_\infty}$, with only a small weight on $\ket{-\alpha_\infty}$. On the other hand, for initial phases $\arg(\alpha_k)=\pm3\pi/4,\pi$, the state mainly evolves to $\ket{-\alpha_\infty}$ with a small weight on $\ket{\alpha_\infty}$. For the initial phases $\arg(\alpha_k)=\pm\pi/2$, the initial state is almost symmetrically positioned with respect to the two states $\ket{\pm\alpha_\infty}$ and has no definite parity (even and odd photon number states are almost equally populated). Hence, the state evolves to a mixture of $\ket{\pm\alpha_\infty}$.}
\label{fig:double_well}
\end{figure*}

\begin{figure*}
\setlength{\unitlength}{1cm}
\includegraphics[width=1.2\columnwidth]{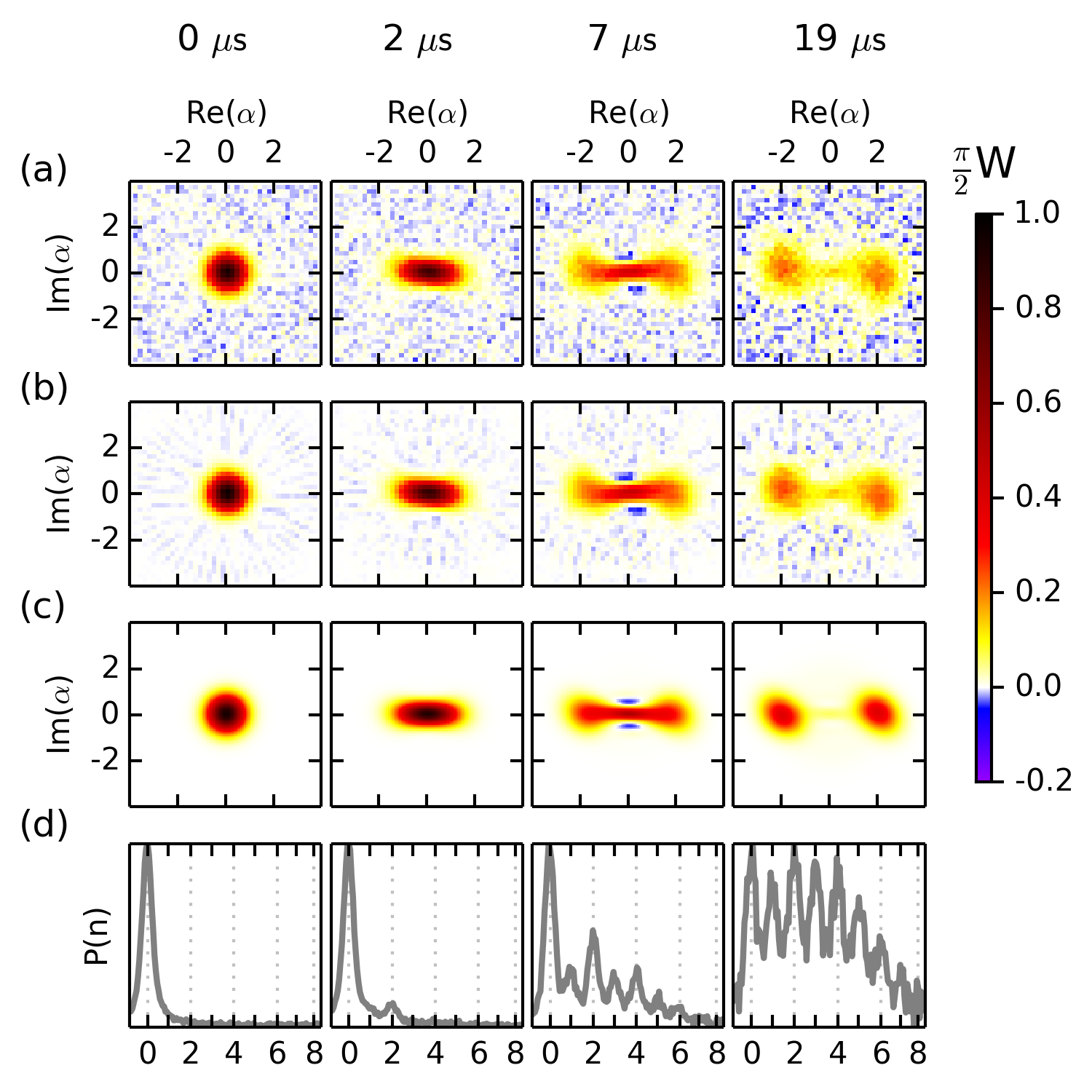}
\caption{Time evolution of the storage state in presence of the nonlinear drive and dissipation processes described in Fig.~\ref{fig:schema}. The panels correspond to measured data (a), to reconstructed density matrices \cite{Vlastakis2013} (b), and to numerical simulations (c). They display the Wigner function after a pumping duration indicated at the top of the panel. The storage is initialized in the quantum vacuum state at $t=0~\mu$s. First, the state squeezes in the $Q$ quadrature ($t=2~\mu$s). Small, but visible negativities appearing at $t=7~\mu$s indicate that the superposition of the SSSs shown in Fig.~\ref{fig:double_well}, panel 2, is now coherent, and that a continuous evolution from a squeezed state to a quantum state approximating a Schr\"{o}dinger cat state is taking place. Finally, these negativities disappear as a consequence of the unavoidable storage photon loss, and the state decays into a statistical mixture of the two SSSs ($t=19~\mu$s). (d) Storage photon number distribution measured using the photon number splitting of the qubit \cite{Schuster2007}. At $t=2,7~\mu$s, the $n=2$ population is larger than $n=1$. A similar population inversion is also present between $n=4$ and $n=3$ at $t=7~\mu$s. The non-Poissonian character of the photon number distribution at $t=2,7~\mu$s confirms the non-classical nature of the dynamical states of the storage for these intermediate times.}
\label{fig:time_evolution}
\end{figure*}




\clearpage

\beginsupplement
\renewcommand{\bibnumfmt}[1]{[S#1]}
\renewcommand{\citenumfont}[1]{S#1}

\widetext
\begin{center}
\textbf{\large Supplementary material for ``Confining the state of light to a quantum manifold by engineered two-photon loss''}
\end{center}

\section{Materials and methods}

\subsection{Qubit fabrication}
The transmon qubit was fabricated with a double-angle-evaporated Al/AlO$_x$/Al Josephson junction, defined using the bridge-free  fabrication technique \cite{Lecocq2011} on a double-side-polished 2 mm-by-19 mm chip of c-plane sapphire with a 0.43 mm thickness. The aluminum film thickness for each deposition was 20 nm and 30 nm. The Josephson junction has {an area of 0.09$\pm$ 0.02 $\mu$m$^2$}. Between these two depositions, an AlO$_x$ barrier was grown via thermal oxidation for {6 minutes} in 100 Torr static pressure of gaseous mixture 85 \% argon 15 \% oxygen. The room-temperature junction resistance was 6.67 k$\Omega$.

The sapphire chip was placed across two 3D aluminum cavities separated by a 2 mm wall, {as shown in Fig.~\ref{fig:cavPic}}. These cavities were machined out of high purity aluminum (99.99\% purity), and prepared by removing $\approx~200~\mu$m of material with acid etching \cite{Reagor2013}. The antenna pads on each side of the Josephson junction couple to the TE101 mode of each cavity. On the readout cavity side, the antenna is 0.5 mm wide and 7.5 mm long. On the storage cavity side, the antenna is 0.5 mm wide and 4.2 mm long with a 0.01 mm gap capacitor for extra coupling tunability. These dimensions were optimized to meet the desired coupling strengths using finite element simulations and black box circuit quantization analysis \cite{Nigg2012}.

\subsection{Measurement setup}
\subsubsection{Waveguide Purcell filter}
The output of the readout cavity is coupled to a transmission line through a WR-102 waveguide which exponentially attenuates signals below a cutoff frequency of 5.8 GHz. This way, the readout (7.152 GHz) is above cutoff, and is hence well coupled to the transmission line. On the other hand, the qubit (4.9007 GHz) is below, and is hence isolated from the transmission line. With this architecture, we obtained a qubit lifetime of $T_1=23~\mu$s despite its strong coupling to the low $Q$ readout cavity ($\chi_{qr}/2\pi=35~$MHz, $\kappa_r=(26~\text{ns})^{-1}$). Waveguide transmission at the qubit frequency is set by the waveguide length (7.62 cm) and detuning below cutoff, and in our case is -70 dB (at 300 K), while only -0.16 dB (at 300 K) at the cavity frequency. The coupling between the cavity and waveguide is through an aperture, whose dimensions (7.4 mm long, 3.96 mm wide, 5.64 mm deep) determine the coupling strength, which is measured to be $Q^{\text{out}}_r$  = 7500 (assuming internal quality factor $Q_r^\text{in}\gg Q^{\text{out}}_r$). The input couplings for the readout and storage cavities were measured at room temperature to be $Q^\text{in}_r$=4,000,000 and $Q^\text{in}_s$=15,000,000. The output port of the storage cavity $Q^\text{out}_s\approx Q^\text{in}_s$ was not used in this experiment.

\subsubsection{Amplification chain}
The transmission line is connected to a Josephson parametric converter (JPC) acting as a phase preserving amplifier \cite{Bergeal2010,Bergeal2010a,Roch2012}, operating near the quantum limit with a gain of 20 dB over a bandwidth of 4.6 MHz. We obtain an input noise visibility ratio for the amplification chain of 8 dB \cite{Narla2014}, indicating that $\approx$90\% percent of the noise at room temperature are amplified quantum fluctuations. The qubit state is measured by sending a square pulse of length $T_\text{pulse}=1~\mu$s through the input readout port. The frequency of this pulse is centered at the readout cavity frequency when the qubit is in its ground state. When the qubit is in its ground state $\ket{g}$, the pulse transmits to the cavity output port towards the JPC. Since the dispersive shift is much larger than the cavity line width ($\chi_{qr}\gg \kappa_r\gg1/T_\text{pulse}$), if on the other hand, the qubit is in its excited state $\ket{e}$, the pulse reflects off the input port. When the qubit is in $\ket{g}$, the steady state number of photons in the readout cavity during this pulse is about 4 photons (calibrated using qubit measurement induced dephasing \cite{Gambetta2006}). 

When exiting the JPC, the pulse propagates through two isolators at 20 mK, a superconducting line between the 20 mK stage and the 4 K stage, where it is amplified by a HEMT amplifier with 40 dB gain. At room temperature, the signal is further amplified, mixed down to 50 MHz and digitized with an analog to digital converter (ADC) (see Fig.~\ref{fig:setup}). For each measurement, we record the two quadratures ($I$ and $Q$) of the digitized signal.  A histogram of 820,000 measured ($I$,$Q$) values is shown in Fig.~\ref{fig:JPC}. This histogram is the sum of two gaussians: the right one corresponds to the qubit in $\ket{g}$ and the left one corresponds to the qubit in $\ket{e}$ (corresponding to a qubit thermal excited state occupancy of 20\%). The $I$ and $Q$ quadratures are rotated such that the information lies in the $I$ quadrature only. The right gaussian is squeezed in the $Q$ quadrature, which is a consequence of  the JPC saturation. An ($I$,$Q$) value lying on the right (left) hand side of the threshold indicated by dotted line in Fig.~\ref{fig:JPC} is associated to a qubit in the ground (excited) state. This threshold is calculated such that the errors of mistaking $\ket{g}$ for $\ket{e}$ and $\ket{e}$ for $\ket{g}$ are equal. This separability fidelity is calculated here to be 99\%, which would coincide with the measurement fidelity in the limit of large $T_1$.

\subsection{System parameters}

\subsubsection{Parameter values}
The system parameters are shown in Table~\ref{table:parameters}.
\begin{table}[!h]
\begin{tabular}{| l | c | c | c | c | }
\hline 
  Mode & Frequency (GHz) & T$_1$  ($\mu$s) & T$_2$  ($\mu$s) & Thermal population\\
  \hline 
  Qubit & $ 4.9007 $ & 23 & 1 & 20\%\\
  Storage & 7.57861 & 20 & - &$\le$ 5\%\\
  Readout & 7.152 & 0.025 & - &$\le$ 2\%\\
  \hline 
\end{tabular}
\caption{Frequencies, thermal populations and coherence times of each mode.}
\label{table:parameters}
\end{table}

\begin{table}[!h]
\begin{tabular}{| l | c | c | c | }
\hline 
  $\chi/2\pi$ (MHz) & Qubit & Storage & Readout\\
  \hline 
  Qubit & $ 130 $ &  & \\
  Storage & 1.585 & (0.004) &  \\
  Readout & 35 & 0.206 &2.14\\
  \hline 
\end{tabular}
\caption{Dispersive couplings between the qubit, storage and readout modes. The diagonal elements in this table refer to the self-Kerr terms, which enter in the Hamiltonian as $\sum_{m=q,r,s}-\frac{\chi_{mm}}{2}{\ba_m^\dag}^2\ba_m^2$, where the subscripts $m=q,r,s$ stand respectively for the qubit, readout and storage. The off-diagonal terms in the table are the cross-Kerr terms, which enter the Hamiltonian as $-\chi_{qs}\ba_q^\dag\ba_q\ba_s^\dag\ba_s-\chi_{qr}\ba_q^\dag\ba_q\ba_r^\dag\ba_r-\chi_{rs}\ba_r^\dag\ba_r\ba_s^\dag\ba_s$. The value for the storage Kerr (between brackets) was not directly measured, but only estimated from other measured quantities using the geometric equality: $\chi_{ss}=\chi_{qs}^2/4\chi_{qq}$ \cite{Nigg2012}.}
\end{table}

\subsubsection{Choice of parameters}
As described in the main text, the goal of this experiment was to obtain a non-linear dissipation rate $\kappa_2=\frac{\chi_{sr}^2}{\kappa_r}\abs{\xi_p}^2$ which is as large as possible. This rate is proportional to the pump power and the square of the readout-storage cross-Kerr $\chi_{sr}$. It is not possible to pump arbitrarily hard since mixing of the pump due to higher order non-linear terms will eventually produce undesirable effects. For example, in Fig.~\ref{fig:Stark}, we can see that for pump powers larger than 100 mW {(measured at the output of the generator)}, the storage mode linewidth increases above the linewidth in absence of pump. We have also seen that for pump powers larger than 200 mW, the qubit thermal population starts to increase. This is why we fix the pump power to 100 mW for the rest of the experiment. From the AC Stark shift on the qubit, we know that this corresponds to $\abs{\xi_p}^2=1.2$. Therefore, it is useful to have a large enough $\chi_{sr}$ in order to achieve $\kappa_2$ of the same order as $\kappa_s$ for $\abs{\xi_p}\approx1$. For our parameter values, this corresponds to $\chi_{rs}/2\pi$ of the order of 200 kHz. We designed our system to obtain the latter coupling. We cannot increase this coupling too much since we believe this will decrease the storage cavity lifetime due to the Purcell effect (in the near future, we plan on designing a pass-band, instead of a high-pass, Purcell filter to lift this constraint). Since we have $\chi_{rs}=2\sqrt{\chi_{rr}\chi_{ss}}$, and we want a storage Kerr at most of the order of its linewidth (in order to minimize the distortion of the coherent state superpositions), we had to increase the readout Kerr $\chi_{rr}$ (by increasing the junction participation in this mode \cite{Nigg2012}) until we obtained the desired $\chi_{rs}$. Moreover, we needed to have a qubit mode to perform Wigner tomography. The latter necessitates short un-selective pulses on the qubit \cite{Vlastakis2013}. For this reason, we needed a large enough transmon anharmonicity, which necessarily implied a very large qubit-readout cross-Kerr (here $\chi_{qr}/2\pi=35~$MHz). 

Strongly coupling a qubit to a lossy resonator reduces its coherence times due to the Purcell effect. The use of a Purcell filter \cite{Reed2010} (described above) seemed favorable. This is why we designed our qubit frequency to be around 5 GHz,  and the readout mode around 7 GHz, the former below and the latter above the waveguide cutoff frequency. The pump tone needs to be at $\omega_p=2\omega_s-\omega_r$, which is below the readout if $\omega_s<\omega_r$ and above otherwise. We thought it would be more cautious to have this strong pump tone as far as possible from the qubit (to avoid the pump coupling to the qubit mode), and therefore designed the storage mode to be about half a GHz above the readout. This way, the pump is one GHz above the readout mode, and hence three GHz above the qubit. The drawback of this design is that the storage mode is not protected by the Purcell filter since it is above cutoff. In the near future we will repeat this experiment with a pass-band Purcell filter.

\subsection{Measurement methods}
\subsubsection{Spectroscopy} 



Readout mode and qubit spectroscopy are obtained by performing transmission spectroscopy and saturation spectroscopy, respectively. Storage mode spectroscopy is obtained by sequentially sending a long ($100~\mu s$)  and weak probe tone to the storage input port, then performing a selective $\pi$ pulse \cite{Johnson2010} on the qubit conditioned on there being zero photons in the storage, and finally measuring the qubit through the readout mode. If the probe tone is off-resonant with the storage mode frequency, the storage photon number remains zero, the $\pi$ pulse therefore inverts the qubit state. On the other hand, if the probe tone is resonant, the storage gets populated to larger photon numbers, hence the $\pi$ pulse cannot completely invert the qubit state. This change in qubit state vs. probe frequency is detected by the measurement pulse through the readout mode.

\subsubsection{Lifetimes}
Qubit lifetime $T_1$ and coherence time $T_2$ are measured with the usual $T_1$ and Ramsey pulse sequences. The readout mode lifetime is extracted from its linewidth. Since the readout mode has a relatively large Kerr ($\chi_{rr}/2\pi=2.14~$MHz), the transmission spectra are broadened by this Kerr as we increase the power of the probe tone. Hence, we perform transmission spectroscopy for decreasing probe power until the linewidth stops narrowing. The mode lifetime is then $1/\kappa_r$ where $\kappa_r/2\pi$ is the spectral linewidth at small probe powers. The storage mode lifetime is obtained by first displacing the storage state, and after a variable wait time, measuring the parity of the storage state. By fitting the data, we obtain the storage lifetime.

\subsubsection{Thermal population}
Qubit thermal population is obtained by taking a single shot histogram of the qubit state (see Fig.~\ref{fig:JPC}). We get the thermal excited state occupancy by extracting the probability of getting a count on the left hand side of the threshold (dotted line). We can give a bound on the thermal population $n^{\text{th}}_r$ of the readout mode. This thermal population $n^{\text{th}}_r$ induces a dephasing rate for the qubit given by $\kappa_{\phi,th}=n^{\text{th}}_r\kappa_r$, in the limit where $\chi_{qr}\gg \kappa_r$ \cite{Sears2012}. We know that the measured dephasing rate $\kappa_\phi=1/T_2-1/2T_1\approx 1/T_2$ (since $T_1\gg T_2$), is at least larger than $\kappa_{\phi,th}$. The inequality $\kappa_\phi\ge\kappa_{\phi,th}$ is equivalent to
$$
n^{\text{th}}_r\le1/(T_2\kappa_r)= 2\%\;.
$$
By measuring the qubit number split spectrum to the storage mode, we should in principle be able to measure the storage thermal occupancy. However, the spectrum linewidth sets a bound on which thermal population in the storage one can robustly measure. This linewidth $\kappa_\text{spec}/2\pi$ is due to the finite spectroscopy pulse length and power, and is bounded by $(2\pi T_2)^{-1}$. Assuming a small number of thermal photons $n^{\text{th}}_s\ll1$, at equilibrium, the storage is in a mixture of the vacuum state with probability $(1-n^{\text{th}}_s)$ and the first excited state with probability $n^{\text{th}}_s$. The spectrum of the qubit is then $S(\omega)=(1-n^{\text{th}}_s)S_0(\omega)+n^{\text{th}}_sS_1(\omega)$, where $S_0$ and $S_1$ are the qubit spectra when the number of photons in the storage is 0 or 1, respectively. We have $S_k(\omega)=\frac{\abs{\epsilon_\text{probe}}^2}{\left(\frac{\kappa_\text{spec}}{2}\right)^2+(\omega-\omega_q-k\chi_{qs})^2}$, where $\epsilon_\text{probe}$ is the probe amplitude, and we have neglected the effect of $\kappa_s$ on $\kappa_\text{spec}$ since in practice $\kappa_s\ll\kappa_\text{spec}$. When we measured the spectrum $S$ while the storage was in thermal equilibrium, we could not resolve a peak at $\omega_q-\chi_{qs}$ corresponding to one photon. This implies that we have $n^{\text{th}}_sS_1(\omega_q-\chi_{qs})\le(1-n^{\text{th}}_s)S_0(\omega_q-\chi_{qs})$.  In our case, we took a qubit spectrum with a gaussian $\pi$ pulse ($800$ ns standard deviation), and we observed a linewidth $\kappa_{\text{spec}}$=1/(0.23~$\mu$s). In the limit where $\kappa_\text{spec}\ll \chi_{qs}$, this sets the following bound on our measure of $n^{\text{th}}_s$ :

$$
n^{\text{th}}_s\le(\kappa_{\text{spec}}/2\chi_{qs})^2 = 5\%\;.
$$ 

\subsubsection{Cross-Kerr terms}
The qubit to readout cross-Kerr is obtained by measuring the readout spectrum. Due to the thermal occupancy of the qubit, this spectrum exhibits two peaks, separated by $\chi_{qr}/2\pi$. The qubit to storage cross-Kerr is obtained by inserting photons in the storage and measuring a qubit spectrum. We see many peaks, each one corresponding to a photon number state in the storage. The linear dependence of the central frequency of each peak on the peak number give the qubit-storage cross-Kerr (see Fig.~\ref{fig:qubitnumbersplitting}). This measurement is further refined by performing a parity revival experiment \cite{Vlastakis2013}. The readout to storage cross-Kerr is obtained by measuring the readout frequency as a function of photons inserted in the storage. The readout mode frequency decreases linearly with storage photon number with a proportionality constant corresponding to the cross-Kerr.

\subsubsection{Kerr terms}
The transmon anharmonicity (also termed qubit Kerr $\chi_{qq}$) is obtained by measuring qubit spectroscopy with increasing probe power until we observe the two photon transition from $\ket{g}$ to $\ket{f}$, which is detuned from the main $\ket{g}$ to $\ket{e}$ peak by half the qubit anharmonicity. The readout mode Kerr is obtained from the pump Stark shift (Fig.~\ref{fig:Stark}). Indeed, as we will show in the following section, due to the pump, all three modes frequencies decrease linearly with the pump power. The ratio of the slopes of the qubit shift to the readout shift is $\chi_{qr}/2\chi_{rr}$. Hence, knowing $\chi_{qr}$, we extract $\chi_{rr}$. A useful check is to make sure that the ratio of slopes of the qubit and storage shifts is indeed $\chi_{qr}/\chi_{rs}$. We find that this value agrees with the independently measured cross-Kerr values with a deviation of $5\%$. The storage Kerr was not measured, but merely estimated from the formula $\chi_{ss}=\chi_{qs}^2/4\chi_{qq}$ \cite{Nigg2012}.

\subsubsection{Photon number calibration}
The storage cavity was displaced using a 20 ns square pulse. Similarly to \cite{Kirchmair2013}, we calibrate the amplitude of this pulse by measuring a cut of the Wigner function of the vacuum state, and fitting a gaussian to the data. The DAC to photon number correspondence is obtained by imposing that the standard deviation of this gaussian needs to be $1/2$. We calibrate the number of photons in the readout mode by measuring the measurement-induced dephasing rate on the qubit while a tone is applied to the readout mode \cite{Gambetta2006}.

\subsubsection{Phase locking}
The quantum state produced in the storage is a consequence of non-linear mixing of the pump and drive tones in our Josephson circuit. If we used a third generator to probe the state of the storage, this generator would not be phase locked to the state in the storage, and hence we would expect all our Wigner functions to be completely smeared and to exhibit no phase coherence. To avoid this problem we generate the pump and storage tones from two separate generators at respectively $\omega_p$ and $\omega_s$, and we mix them at room temperature to generate the drive tone (see dashed box in Fig.~\ref{fig:setup}). This is achieved by doubling the frequency of the storage generator to $2\omega_s$ using a mixer, and then mixing this doubled frequency with the pump to obtain $2\omega_s\pm\omega_p$. The upper sideband at $2\omega_s+\omega_p$ is then filtered by a low pass filter with a 12 GHz cutoff frequency, and hence only the drive tone at the desired frequency $\omega_d=2\omega_s-\omega_p$ enters our device. We use fast microwave switches controlled by markers from the arbitrary waveform generator (AWG) to produce the pulse sequences for the experiment.

\subsubsection{Parity measurement and Wigner tomography}
\label{sec:WignerTomo}

The Wigner function uniquely defines the quantum state $\rho_s$ of an oscillator. It is defined as $W(\alpha)=\frac{2}{\pi}P(\alpha)$, where $P(\alpha)=\tr{\bD_{-\alpha}\rho_s\bD_{\alpha}e^{i\pi\ba_s^\dag\ba_s}}$ \cite{Haroche2006}.

In this experiment, we directly measured $P(\alpha=I+iQ)$ following the measurement protocol of \cite{Vlastakis2013,Sun2014} (see Fig.~\ref{fig:pulseSeq}). In the data of Figs.~\ref{fig:time_evolution}-\ref{fig:Fock_time_evolution}, for each point $(I_k,Q_k)$ of the $I-Q$ plane, we repeat 10,000 times:
\begin{enumerate}
\item Initialize the qubit by measuring its state and post-selecting on it being in the ground state
\item Displace the cavity state with a 20 ns square pulse of amplitude $a_k=\sqrt{I_k^2+Q_k^2}$ and phase $\phi_k=\arg(I_k+iQ_k)$
\item Perform a $+\pi/2$ pulse on the qubit around the X-axis.
\item Wait for $\pi/\chi_{qs}$
\item Perform a $+\pi/2$ pulse on the qubit around the X-axis (then repeat all steps with a $-\pi/2$ pulse)
\item Measure the qubit state
\end{enumerate}

All measurements are single shot and are binned to be 0 or 1 depending on whether the data point lies on the left or right of the threshold (see Fig.~\ref{fig:JPC}). Each one of the qubit pulses is a gaussian pulse with a 4 ns standard deviation, and we truncate the pulse length to 5 standard deviations. After post-selecting on the initial measurement, the data is averaged, and two Wigner maps are obtained. One corresponding to both pulses with a $+\pi/2$ angle, and the other where the second pulse is with a $-\pi/2$ angle. We then subtract these two maps in order to correct for systematic errors due to the readout-storage cross-Kerr and the finite un-selectivity of the $\pi/2$ pulses  \cite{Vlastakis2013,Sun2014}.

{Indeed, assume the storage is in a pure state $\ket{\psi}$, and we want to measure its Wigner function. We model the finite un-selectivity of the $\pi/2$ pulses by assuming that there is an $N_{\text{max}}$, such that if there are $n\le N_\text{max}$ photons in the cavity, the pulses are able to rotate the qubit state, whereas for all $n> N_\text{max}$, the qubit state is unaffected by the pulse. Each qubit measurement is thresholded and associated to the qubit being in state g or e. The probability of measuring $m=g,e$ when the qubit state was in fact in $t=g,e$ is denoted $p^\alpha(m|t)$. In the latter notation, the superscript $\alpha$ refers to the displacement amplitude of the storage, which is a simplified model incorporating the readout-storage cross-Kerr and its effect on the readout fidelity due to the presence of photons in the storage. First, we displace the state by $\alpha$, and denote the displaced state $\ket{\psi_\alpha}$. Second, we perform two $\pi/2$ pulses separated by a $\pi/\chi_{qs}$ wait time. We then obtain the following qubit-storage entangled state : 
$$\ket{\psi_\alpha}^+=\bP_\text{even}\ket{\psi_\alpha}\ket{e}+\bP_\text{odd}\ket{\psi_\alpha}\ket{g}+\bP_{>N_\text{max}}\ket{\psi_\alpha}\ket{g}\;,$$
where $\bP_\text{even}=\sum_{2n\le N_\text{max}}{\ket{2n}\bra{2n}}$, $\bP_\text{odd}=\sum_{2n+1\le N_\text{max}}{\ket{2n+1}\bra{2n+1}}$ and $\bP_{>N_\text{max}}=\sum_{n > N_\text{max}}{\ket{n}\bra{n}}$. The measured quantity, which is the expectation value of the qubit energy is 
\begin{eqnarray*}
\bket{\sigma_z}^+&=&\norm{\bP_\text{even}\ket{\psi_\alpha}}^2(p^\alpha(e|e)-p^\alpha(g|e))-\norm{\bP_\text{odd}\ket{\psi_\alpha}}^2(p^\alpha(g|g)-p^\alpha(e|g))\\
&+&\norm{\bP_{>N_\text{max}}\ket{\psi_\alpha}}^2 (p^\alpha(e|g)-p^\alpha(g|g))\;.
\end{eqnarray*}
When the second $\pi/2$ pulse has a $\pi$ phase shift, we get 
$$\ket{\psi_\alpha}^-=\bP_\text{even}\ket{\psi_\alpha}\ket{g}+\bP_\text{odd}\ket{\psi_\alpha}\ket{e}+\bP_{>N_\text{max}}\ket{\psi_\alpha}\ket{g}\;,$$
and hence 
\begin{eqnarray*}
\bket{\sigma_z}^-&=&\norm{\bP_\text{even}\ket{\psi_\alpha}}^2(p^\alpha(e|g)-p^\alpha(g|g))-\norm{\bP_\text{odd}\ket{\psi_\alpha}}^2(p^\alpha(g|e)-p^\alpha(e|e))\\
&+&\norm{\bP_{>N_\text{max}}\ket{\psi_\alpha}}^2 (p^\alpha(e|g)-p^\alpha(g|g))\;.
\end{eqnarray*}
We then substract these two expectation values and obtain $\Delta\bket{\sigma_z}=C_\alpha(\norm{\bP_\text{even}\ket{\psi_\alpha}}^2-\norm{\bP_\text{odd}\ket{\psi_\alpha}}^2)=C_\alpha P(\alpha)$, where the contrast $C_\alpha$ is given by $C_\alpha=\frac{1}{2}(p^\alpha(g|g)+p^\alpha(e|e)-p^\alpha(e|g)-p^\alpha(g|e))$. In the case of perfect readout: $p^\alpha(g|g)=p^\alpha(e|e)=1$ and $p^\alpha(e|g)=p^\alpha(g|e)=0$, and hence $C_\alpha= 1$.  Notice that this subtraction eliminated the third term in $\bket{\sigma_z}^\pm$ which is due to the finite un-selectivity of the pulses, and would appear as an offset in the Wigner tomography. This subtraction also makes the effect of the storage-readout cross-Kerr symmetric, making no bias towards positive or negative values.}

{From these measured Wigner functions, one can reconstruct a density matrix which best reproduces this data \cite{Vlastakis2013}. As a consistency check, we can compare the diagonal elements of this reconstructed density matrix, to the directly measured photon number probabilities using qubit spectroscopy. As shown in Fig.~\ref{fig:Numbersplitting_time_evolution}, there is a good agreement between these two independent measurements. One can also extract the expectation value of any observable directly from the measured Wigner function, and compare them to the theoretical predictions through numerical simulations. This comparison is made in Figs.~\ref{fig:Obs_vs_time}-\ref{fig:Obs_vs_time_2}, and we observe good agreement between theory and experiment.}

\subsubsection{Qubit dynamics during the pumping}
When the pump and the drive tones are on, the readout mode remains mainly in vacuum and the storage state evolves from vacuum to a mixture of coherent states, while transiting through a coherent state superposition (see Fig.~4 of the main text). In principle, if the Hamiltonian of the three modes (qubit, readout, storage) is fully captured by the Hamiltonian described in \eqref{eq:totalH1}\eqref{eq:totalH2}, the qubit state should not be influenced by the pumping. For example, if we initialize the qubit in its ground state before activating the pump and drive tones, the qubit should remain in its ground state, unless it absorbs a thermal photon, and this thermal absorption rate should be independent of the number of photons in the two other modes. However, we have observed that when the pumping is on, as the photon number in the storage mode increases, the qubit thermal occupation increases significantly. This is most likely related to the previously unexplained mechanism which causes the qubit lifetime to decrease when photons are inserted in the readout mode \cite{Slichter2012}.

The parametric pumping mechanism relies on the frequency matching condition $\omega_p=2\omega_s-\omega_r$, where $\omega_{p,s,r}$ are the pump, storage and readout frequencies, respectively. The pump and drive tone frequencies need to be tuned with a precision of order $g_2$, as observed in Fig.~2 of the main paper, and computed in the next section (see Eq.~\eqref{eq:g2}). In our experiment, we tune these tones to fulfill this condition when the qubit is in its ground state. If the qubit suddenly jumps to the excited while the pumping is activated, the readout and storage frequencies will shift by their respective dispersive coupling to the qubit $\chi_{qr}$ and $\chi_{qs}$. In particular, $(\chi_{qr}-\chi_{qs})/2\pi=33.4~$MHz $\gg g_2/2\pi=111~$kHz. Hence, the frequency matching condition no longer holds, and the pumping process is interrupted. This undesirable process can be slightly filtered by measuring the qubit state after completing the pumping, and post-selecting on the qubit being in its ground state (see Fig.~\ref{fig:pulseSeq}). However, we do not filter out processes where the qubit jumped up to the excited state for a random time, and jumped back down to its ground state before the measurement is performed. We believe it is these kinds of processes which produce an excess of n=0 population in the storage (see Fig.~\ref{fig:Numbersplitting_time_evolution}). The effect of these large pumps and populated modes on the qubit decay rates is subject to ongoing research.

\section{Supplementary text}

\subsection{The pumped Josephson circuit Hamiltonian}
We start by writing the Hamiltonian of the qubit, readout and storage modes coupled to a Josephson junction, with two tones (the drive and the pump) on the readout mode.
\begin{eqnarray*}
\bH/\hbar&=&\sum_{m=q,r,s}\bar{\omega}_m\ba_m^\dag\ba_m-\frac{E_J}{\hbar}\left(\cos(\bphi)+\bphi^2/2\right)
+2\Re\left(\epsilon_pe^{-i\omega_p t}+\epsilon_de^{-i\omega_d t}\right)(\ba_r+\ba_r^\dag)\;,\\
\bphi&=&\sum_{m=q,r,s}\varphi_m(\ba_m+\ba_m^\dag)\;.
\end{eqnarray*}
The first term corresponds to the linear Hamiltonian of each mode of annihilation operator $\ba_m$. Their bare frequencies $\bar\omega_m$ are shifted towards the measured frequencies $\omega_m$ due to the contribution of the Josephson junction in the Hamiltonian. The latter is represented by the cosine term, to which we have removed the quadratic terms by including them in the linear part of the Hamiltonian. $E_J$ is the Josephson energy, and $\bphi$ is the phase across the junction, which can be decomposed as the linear combination of the phase across each mode, with $\varphi_m$ denoting the contribution of mode $m$ to the zero point fluctuations of $\bphi$. The system is irradiated by a drive and pump tones with complex amplitudes $\epsilon_d,~\epsilon_p$ and frequencies $\omega_d,~\omega_p$, respectively. $\Re()$ denotes the real part. The pump is a large amplitude far off-resonant tone, while the drive is a weak tone close to resonant with the readout mode.

We place ourselves in a regime where 
$$
\omega_{p},~\omega_d,~\bar\omega_m \gg \epsilon_p\sim (\omega_p-\bar\omega_r)\gg \frac{E_J}{\hbar} \norm{\bphi}^4/4!\;.
$$

In order to eliminate the fastest time scales corresponding to the system frequencies and the pump amplitude, we make a change of frame using the unitary 
\begin{eqnarray*}
U&=&{e^{i\bar\omega_q t \ba_q^\dag\ba_q}}{e^{i\omega_d t \ba_r^\dag\ba_r}}{e^{i\frac{\omega_p+\omega_d}{2} t \ba_s^\dag\ba_s}}e^{-\tilde\xi_p\ba_r^\dag+\tilde\xi_p^*\ba_r}\;,\\
\frac{d\tilde\xi_p}{dt}&=&-i\bar\omega_r\tilde\xi_p-i2\Re\left(\epsilon_pe^{-i\omega_pt}\right)-\frac{\kappa_r}{2}\tilde\xi_p\;.
\end{eqnarray*}
After a time scale of order $1/\kappa_r$ we have $\tilde\xi_p\approx \xi_pe^{-i\omega_p t}$, $\xi_p={-i\epsilon_p}/\left({\frac{\kappa_r}{2}+i(\bar\omega_r-\omega_p)}\right)\approx{-i\epsilon_p}/\left({\frac{\kappa_r}{2}+i(\omega_r-\omega_p)}\right)$.

In this new frame, the Hamiltonian is
\begin{eqnarray*}
\tilde\bH/\hbar&=&(\bar\omega_r-\omega_d)\ba_r^\dag\ba_r+(\bar\omega_s-\frac{\omega_p+\omega_d}{2})\ba_s^\dag\ba_s-\frac{E_J}{\hbar}(\cos(\tilde\bphi)+\tilde\bphi^2/2)\;,\\
\tilde\bphi&=&\sum_{k=q,r,s}\phi_k(\tilde\ba_k+\tilde\ba_k^\dag)+(\tilde\xi_p+\tilde\xi_p^*)\phi_r\;,\\
\tilde\ba_q&=&e^{-i\bar\omega_q t}\ba_q\;,\tilde\ba_r=e^{-i\omega_d t}\ba_r\;, \tilde\ba_s=e^{-i\frac{\omega_p+\omega_d}{2} t}\ba_s\;.
\end{eqnarray*}

We now expand the cosine up to the fourth order, and only keep non rotating terms:
\begin{eqnarray}
\tilde\bH&\approx&\bH_{\text{shift}}+\bH_{\text{Kerr}}+\bH_{2}\;,
\label{eq:totalH1}
\end{eqnarray}
where : 
\begin{eqnarray}
\bH_{\text{shift}}&=&(-\delta_q-\chi_{qr}\abs{\xi_p}^2)\ba_q^\dag\ba_q\notag\\
&+&(\bar\omega_r-\omega_d-\delta_r-{2\chi_{rr}}\abs{\xi_p}^2)\ba_r^\dag\ba_r\notag\\
&+&(\bar\omega_s-\frac{\omega_p+\omega_d}{2}-\delta_s-\chi_{rs}\abs{\xi_p}^2)\ba_s^\dag\ba_s\;,\notag\\
\bH_{\text{Kerr}}&=&-\sum_{m=q,r,s}\frac{\chi_{mm}}{2}{\ba_m^\dag}^2\ba_m^2-\chi_{qr}\ba_q^\dag\ba_q\ba_r^\dag\ba_r-\chi_{qs}\ba_q^\dag\ba_q\ba_s^\dag\ba_s-\chi_{rs}\ba_r^\dag\ba_r\ba_s^\dag\ba_s\;,\notag\\
\bH_{2}&=&g_{2}^*\ba_s^2\ba_r^\dag+g_{2}(\ba_s^\dag)^2\ba_r+\epsilon_d\ba_r^\dag+\epsilon_d^*\ba_r\;.
\label{eq:totalH2}
\end{eqnarray}
The first term $\bH_{\text{shift}}$ corresponds to the modes frequency shifts. The bare frequencies are shifted by $\delta_{q,r,s}$ which arise from the operator ordering chosen in $\bH_{\text{Kerr}}$. Moreover, the frequencies are shifted down by a term proportional to $\abs{\xi_p}^2$, which corresponds to the AC Stark shift induced by the pump. We observe this linear shift vs. pump power in Fig.~\ref{fig:Stark}.

The second term $\bH_{\text{Kerr}}$ corresponds to self-Kerr and cross-Kerr coupling terms \cite{Nigg2012}. We have: $\chi_{mm}=\frac{E_J}{\hbar}\varphi_m^4/2$, and $\chi_{mm'}=\frac{E_J}{\hbar}\varphi_m^2\varphi_{m'}^2$.

The last term $\bH_2$ contains the terms which reveal the physics we have observed in this paper. It is the microscopic Hamiltonian of a degenerate parametric oscillator \cite{Wolinsky1988}. The first term in this Hamiltonian is a non-linear coupling between the storage and readout modes: two photons from the storage can swap with a single photon in the readout. In contrast to the usual parametric oscillator, our readout mode is not twice the frequency of the storage mode. This term is produced by four-wave mixing of the pump and the readout and storage modes. The term in $\epsilon_d$ corresponds to a drive on the readout mode. Our coupling strength is given by
$$
g_{2}=\chi_{sr}\xi_p^*/2\;.
$$
The second term in $\bH_2$ is a coherent drive on the readout mode. It corresponds to the input energy which is converted into pairs of photons in the storage, thus creating coherent state superpositions.

\subsection{Two-mode model and semi-classical analysis}

Here we assume the qubit remains in its ground state. The storage and readout modes evolve under the Hamiltonian:
\begin{eqnarray}
\bH_{sr}&=&\Delta_d\ba_r^\dag\ba_r+\frac{\Delta_p+\Delta_d}{2}\ba_s^\dag\ba_s\notag\\
&+&g_{2}^*\ba_s^2\ba_r^\dag+g_{2}(\ba_s^\dag)^2\ba_r+\epsilon_d\ba_r^\dag+\epsilon_d^*\ba_r\\
&-&\chi_{rs}\ba_r^\dag\ba_r\ba_s^\dag\ba_s-\sum_{m=r,s}\frac{\chi_{mm}}{2}{\ba_m^\dag}^2\ba_m^2\;,
\label{eq:Hsr}
\end{eqnarray}
where $\Delta_d=\bar\omega_r-\omega_d-\delta_r-{2\chi_{rr}}\abs{\xi_p}^2$ and $\Delta_p=-\Delta_d+2(\bar\omega_s-\frac{\omega_p+\omega_d}{2}-\delta_s-\chi_{rs}\abs{\xi_p}^2)$.
Theory curves of Fig.2 in the main paper are obtained by numerically finding the steady state density matrix of the Lindblad equation with damping operators $\sqrt{\kappa_r}\ba_r$ and $\sqrt{\kappa_s}\ba_s$ and Hamiltonian $\bH_{sr}$.

We now write the quantum Langevin equations with damping, which require including incoming bath fields $\ba_r^{in}$ and $\ba_s^{in}$ \cite{Drummond2010}:

\begin{eqnarray*}
\frac{d}{dt}\ba_r&=&-i[\ba_r,\bH_{sr}]-\frac{\kappa_r}{2}\ba_r+\sqrt{\kappa_r}\ba_r^{in}\;,\\
\frac{d}{dt}\ba_s&=&-i[\ba_s,\bH_{sr}]-\frac{\kappa_s}{2}\ba_s+\sqrt{\kappa_s}\ba_s^{in}\;.
\end{eqnarray*}

The remainder of this section is devoted to gaining some insight into the steady state solutions of the equations above. We simplify this task by neglecting the Kerr terms in the Hamiltonian. This leads us to:
\begin{eqnarray}
\frac{d}{dt}\ba_r&=&-i\Delta_d\ba_r-ig_{2}^*\ba_s^2-i\epsilon_d-\frac{\kappa_r}{2}\ba_r+\sqrt{\kappa_r}\ba_r^{in}\label{Langevin2modes_r}\;,\\
\frac{d}{dt}\ba_s&=&-i\frac{\Delta_p+\Delta_d}{2}\ba_s-2ig_{2}\ba_s^\dag\ba_r-\frac{\kappa_s}{2}\ba_s+\sqrt{\kappa_s}\ba_s^{in}\;. \label{Langevin2modes_s}
\end{eqnarray}

We can further simplify these nonlinear Langevin equations by taking the classical limit, where the field operators are replaced by their complex expectation values \cite[chapter 4]{Drummond2010}:
\begin{eqnarray*}
0&=&-i\Delta_da_r-ig_{2}^*a_s^2-i\epsilon_d-\frac{\kappa_r}{2}a_r\label{Langevin2modes_r}\;,\\
0&=&-i\frac{\Delta_p+\Delta_d}{2}a_s-2ig_{2}a_s^*a_r-\frac{\kappa_s}{2}a_s \label{Langevin2modes_s}\;.
\end{eqnarray*}

One solution is 
\begin{equation}
a_s=0 \;, a_r=-i\epsilon_d/(\frac{\kappa_r}{2}+i\Delta_d)\;.
\label{eq:g2}
\end{equation}
This is the usual classical Lorentzian response of a driven-damped oscillator. Now assuming $a_s\ne 0$, we obtain a second solution for $a_r$:
$$
a_r=\frac{-\Delta_p-\Delta_d+i\kappa_s}{4g_{2}}e^{2i\theta_s}\;,
$$
where $\theta_s$ is the phase of $a_s$. Here, the modulus squared of $a_r$ is a parabolic function of the detuning $\Delta_p+\Delta_d$ with a width of $1/\abs{4g_2}^2$, and a minimal value $\abs{\kappa_s/4g_2}^2$. This corresponds to the dip observed in Fig.~2 of the main paper, and its depth is a direct signature of the fact that $g_2\gg \kappa_s$. The response of the storage cavity $a_s$ verifies:
\begin{eqnarray*}
a_s^2&=&\frac{1}{g_{2}^*}(-\Delta_d+i\frac{\kappa_r}{2}) a_r-\frac{\epsilon_d}{g_{2}^*}\label{Langevin2modes_r}\;,\\
\abs{a_s}^2&=&\frac{1}{4\abs{g_{2}}^2}(\Delta_d-i\frac{\kappa_r}{2})(\Delta_p+\Delta_d-i\kappa_s)-\frac{\epsilon_d}{g_{2}^*}e^{-2i\theta_s}\;.
\end{eqnarray*}

A sufficient condition for this equation to have a solution is
$$
\frac{\abs{\Delta_d-i\frac{\kappa_r}{2}}\abs{\Delta_p+\Delta_d-i\kappa_s}}{4\abs{g_{2}\epsilon_d}}\le 1\;.
$$

{A model for the response $\abs{a_r}^2$ of the readout mode as a function of the readout probe and pump tone detunings is:}
$$
\abs{a_r}^2(\Delta_r,\Delta_p)=\text{min}\left(\frac{\abs{\epsilon_d}^2}{\frac{\kappa_r^2}{4}+\Delta_d^2},\frac{(\Delta_p+\Delta_d)^2+\kappa_s^2}{16\abs{g_{2}}^2}\right)\;.
$$

We have checked that this simple semi-classical expression without Kerr terms captures the main features of the data in Fig.~2 (a) of the main paper. However, the transient coherent state superposition shown in Fig.~4 of the main paper cannot be explained by such a semi-classical model: it is a quantum signature of our system.

\subsection{Single-mode model and classical analysis}

\subsubsection{Adiabatic elimination of the readout mode}
{We can adiabatically eliminate the readout mode  \cite{Carmichael2007}, and obtain a master equation for the reduced density matrix of the storage mode alone. Let $\rho_{sr}$ be the density matrix which represents the joint readout and storage state. It verifies}
\begin{equation}
\frac{d}{dt}\rho_{sr}=-i[\bH_{sr},\rho_{sr}]+\frac{\kappa_r}{2}\bD[\ba_r]\rho_{sr}+\frac{\kappa_s}{2}\bD[\ba_s]\rho_{sr}\;,
\label{eq:Lindblad_sr}
\end{equation}
{where the Hamiltonian $\bH_{sr}$ is given in \eqref{eq:Hsr}, and here we take $\Delta_d=\Delta_p=0$. Let $\delta$ be a small dimensionless parameter $\delta\ll1$. We place ourselves in the regime where $g_2/\kappa_r, \epsilon_d/ \kappa_r, \chi_{rs}/\kappa_r\sim \delta$ and $\chi_{ss}/\kappa_r,\kappa_s/\kappa_r\sim\delta^2$. We assume that the number of photons in the readout mode is always much smaller than one. We then search for a solution of  \eqref{eq:Lindblad_sr} in the form}
$$
\rho_{sr}=\rho_{00}\ket{0}\bra{0}+\delta\left(\rho_{01}\ket{0}\bra{1}+\rho_{10}\ket{1}\bra{0}\right)+\delta^2\left(\rho_{11}\ket{1}\bra{1}+\rho_{02}\ket{0}\bra{2}+\rho_{20}\ket{2}\bra{0}\right)+O(\delta^3)\;,
$$	
{where $\rho_{mn}$ acts on the storage Hilbert space, whereas $\ket{m}\bra{n}$ act on the readout Hilbert space. The goal here is to derive the dynamics of $\rho_s=\text{Tr}_r(\rho_{sr})=\rho_{00}+\delta^2\rho_{11}$ up to second order in $\delta$, where $\text{Tr}_r$ denotes the partial trace over the readout degrees of freedom. First, lets multiply \eqref{eq:Lindblad_sr} by $\bra{0}$ and $\ket{0}$. We get, up to second order terms in $\delta$ : }

\begin{eqnarray}
\frac{d}{\kappa_r dt}\rho_{00}&=&-\frac{i}{\kappa_r}\bra{0}[\bH_{sr},\rho]\ket{0}+\delta^2\rho_{11}+\frac{\kappa_s}{2\kappa_r}\bD[\ba_s]\rho_{00}+O(\delta^3)\notag\\
&=&-i\delta^2\left(\bA^\dag\rho_{10}-\rho_{01}\bA\right)-i[-\frac{\chi_{ss}}{2\kappa_r}(\ba_s^\dag)^2\ba_s^2,\rho_{00}]+\delta^2\rho_{11}+\frac{\kappa_s}{2\kappa_r}\bD[\ba_s]\rho_{00}\label{eq:lindblad00}\\
&+&O(\delta^3)\notag\;,
\end{eqnarray}
{where $\bA=\frac{1}{\delta\kappa_r}(g_2^*\ba_s^2+\epsilon_d)$, and hence $\norm{A}=O(1)$ in $\delta$.
We now need to find expressions of $\rho_{01,10,11}$ up to $0^{th}$ order terms in $\delta$. We find, neglecting terms of order $\delta$ and higher:}
 \begin{eqnarray}
\frac{d}{\kappa_rdt}\rho_{10}&=&-i\bA\rho_{00}-\frac{1}{2}\rho_{10}+O(\delta)\;,\\
\frac{d}{\kappa_rdt}\rho_{11}&=&-i\left(\bA\rho_{01}-\rho_{10}\bA^\dag\right)-\rho_{11}+O(\delta)\;.
\end{eqnarray}
{The derivative of $\rho_{10}$ has two terms: the first one can be interpreted as an external driving term, and the second is a damping term. Although the first term is time dependent, making this equation difficult to solve exactly, we know that its temporal variation is slow (of order $\delta^2$) in comparison to the damping rate (of order 1). This is where we make the adiabatic approximation: we assume that $\rho_{10}$ is continuously in its steady state. The same reasoning then applies to $\rho_{11}$, which yields: }

 \begin{eqnarray}
\rho_{10}&=&-2i\bA\rho_{00}+O(\delta)\;,\\
\rho_{11}&=&-i\left(\bA\rho_{01}-\rho_{10}\bA^\dag\right)+O(\delta)\\
&=&4\bA\rho_{00}\bA^\dag+O(\delta)\;.
\end{eqnarray}
{
Injecting these expressions in \eqref{eq:lindblad00}, and rearranging terms, we find}

\begin{eqnarray*}
\frac{d}{dt}\rho_s&=&-i[\bH_s,\rho_s]+\frac{\kappa_2}{2}D[\ba_s^2]\rho_s+\frac{\kappa_s}{2}D[\ba_s]\rho_s\;,\\
\bH_s&=&\epsilon_2^*\ba_s^2+\epsilon_2(\ba_s^\dag)^2-\frac{\chi_{ss}}{2}{\ba_s^\dag}^2\ba_s^2\;,
\end{eqnarray*}
{with}
$$
\kappa_2=4\abs{g_2}^2/\kappa_r\;,\qquad \epsilon_2=-2ig_2\epsilon_d/\kappa_r\;.
$$

\subsubsection{Semi-classical analysis}

Let's define $\alpha(t)=\tr{\ba_s\rho_s}$ and calculate its dynamics. Using $[\ba_s,(\ba_s^\dag)^2]=2\ba_s^\dag$, $[\ba_s,{\ba_s^\dag}^2\ba_s^2]=2\ba_s^\dag\ba_s^2$ and $\tr{\ba_s D[\ba_s^2]\rho_s}=-2\tr{\ba_s^\dag\ba_s^2\rho_s}$, we find
\begin{eqnarray*}
\frac{d}{dt}\alpha&=&-2i\epsilon_2\tr{\ba_s^\dag\rho_s}+i\chi_{ss}\tr{\ba_s^\dag\ba_s^2\rho_s}  -\kappa_2\tr{\ba_s^\dag\ba_s^2\rho_s}-\frac{\kappa_s}{2}\alpha\;.
\end{eqnarray*}

Let's assume a solution in the form of a coherent state $\rho_s(t)=\ket{\alpha(t)}\bra{\alpha(t)}$, we then find
\begin{eqnarray*}
\frac{d}{dt}\alpha&=&-2i\epsilon_2\alpha^*-\left(-i\chi_{ss}+\kappa_2\right)\abs{\alpha}^2\alpha-\frac{\kappa_s}{2}\alpha\;.
\end{eqnarray*}
The central panel of Fig.~3 of the main paper illustrates this equation. The white lines correspond to trajectories governed by this equation, and the absolute value $\abs{\frac{d}{dt}\alpha}$ is represented by the colormap.

In steady state $\alpha(t)\rightarrow\alpha_\infty$, and we have
\begin{eqnarray*}
0&=&-2i\epsilon_2\alpha_\infty^*-\left(-i\chi_{ss}+\kappa_2\right)\abs{\alpha_\infty}^2\alpha_\infty-\frac{\kappa_s}{2}\alpha_\infty\;.
\end{eqnarray*}
We write $\alpha_\infty$ in the form $\alpha_\infty=r_\infty e^{i\theta_\infty}$ and $-i\chi_{ss}+\kappa_2=r_2e^{i\varphi_2}$:
\begin{eqnarray*}
2i\epsilon_2r_\infty e^{-i\theta_\infty}&=&-r_2e^{i\varphi_2}r_\infty^2r_\infty e^{i\theta_\infty}-\frac{\kappa_s}{2}r_\infty e^{i\theta_\infty}\;.
\end{eqnarray*}
Notice that $\alpha_\infty=0$ is a solution, now assume $\alpha_\infty\ne 0$:
\begin{eqnarray*}
-2i\epsilon_2e^{-2i\theta_\infty}&=&r_2e^{i\varphi_2}r_\infty^2+\frac{\kappa_s}{2}\;.
\end{eqnarray*}
Taking the module square of this equation we get
\begin{eqnarray*}
r_2^2r_{\infty}^4+r_2\kappa_s\cos(\varphi_2)r_{\infty}^2+\frac{\kappa_s^2}{4}-4\abs{\epsilon_2}^2&=&0\;.
\end{eqnarray*}
The latter equation is quadratic in $r_\infty^2$, and we assume $\varphi_2$ small enough in order for its discriminant to be positive. If $\abs{\epsilon_2}\le\frac{\kappa_s}{4}$, this equation has no positive roots and hence $\alpha_\infty=0$ is the unique solution.

Now lets assume $\abs{\epsilon_2}>\frac{\kappa_s}{4}$, then 
\begin{eqnarray*}
r_{\infty}^2&=&\frac{1}{2r_2^2}\left(-r_2\kappa_s\cos(\varphi_2)+\sqrt{\left(r_2\kappa_s\cos(\varphi_2)\right)^2-4r_2^2(\frac{\kappa_s^2}{4}-4\abs{\epsilon_2}^2)}\right)\;,
\end{eqnarray*}

and two solutions exist for the phase $\theta_\infty$:
\begin{eqnarray*}
\theta_\infty^-&=&\theta_{2}/2+3\pi/4-\varphi_K/2\\
\theta_\infty^+&=&\theta_\infty^-+\pi\;,
\end{eqnarray*}
where $\theta_{2}$ is the phase of $\epsilon_2$, and $\varphi_K=\arctan(\frac{r_{\infty}^2r_2\sin(\varphi_2)}{r_{\infty}^2r_2\cos(\varphi_2)+\kappa_s/2})$

Note that if $\chi_{ss}=0$, then $r_2=\kappa_2$ and $\varphi_2=0$ and we find
$$r_\infty\Big|_{\chi_{ss}=0}=\sqrt{\frac{2\abs{\epsilon_2}-{\kappa_s}/{2}}{\kappa_2}}\;.$$



\clearpage

\begin{figure}[ht!]
\setlength{\unitlength}{1cm}
\includegraphics[width=\columnwidth]{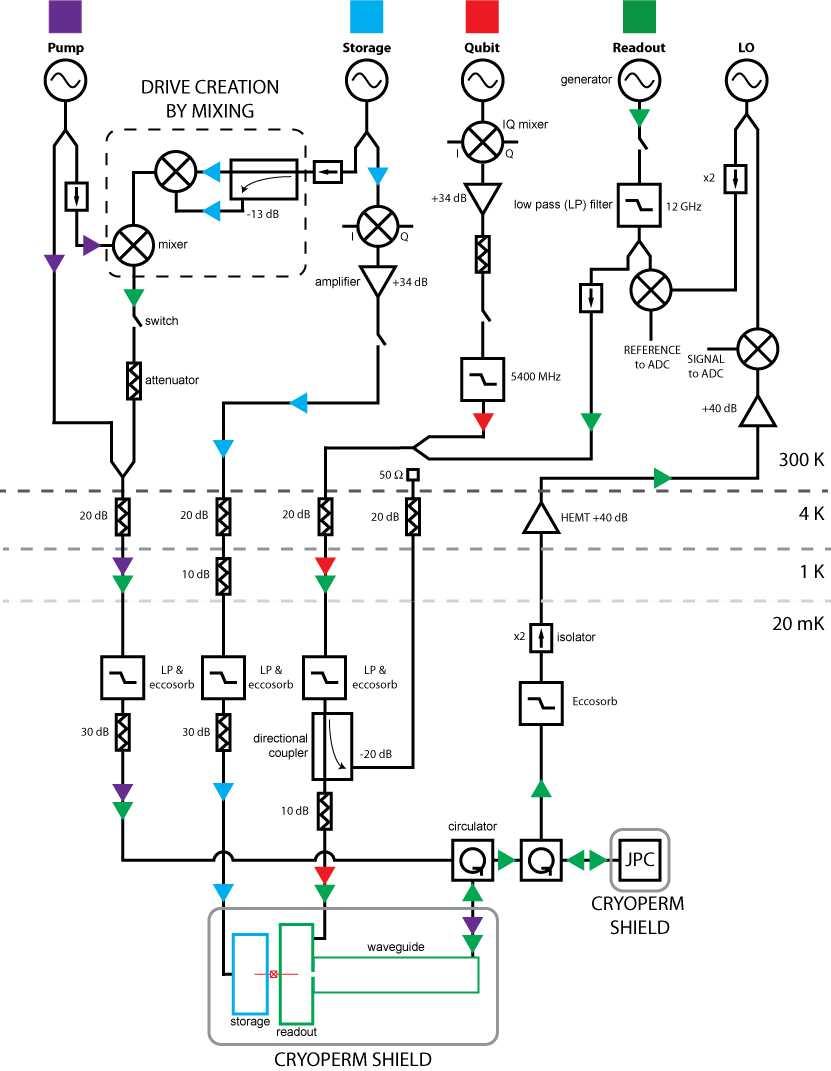}
\caption{Experiment schematic.}
\label{fig:setup}
\end{figure}

\clearpage

\begin{figure}[ht!]
\setlength{\unitlength}{1cm}
\includegraphics[width=0.9\columnwidth]{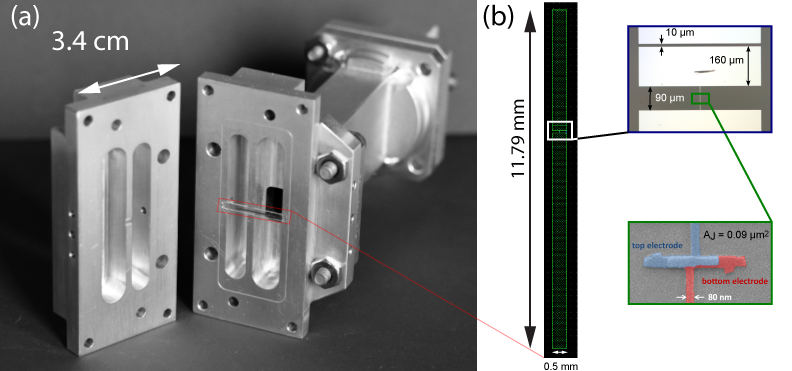}
\caption{{Pictures of the device. (a) Photograph of the two halves of our 3D aluminum cavities, the bridge transmon on a sapphire chip, and the rectangular waveguide. The left half is screwed on to the right one. The readout cavity has a hole which couples it to the rectangular waveguide behind it, which in turn is coupled to a transmission line through a waveguide to SMA adapter. (b) Left : schematics of the JJ and the antenna pads. Top right: optical image in the region containing the JJ and the gap capacitor. Bottom right:  Scanning electron microscope image of the JJ.}}
\label{fig:cavPic}
\end{figure}

\clearpage

\begin{figure}[ht!]
\setlength{\unitlength}{1cm}
\includegraphics[width=0.8\columnwidth]{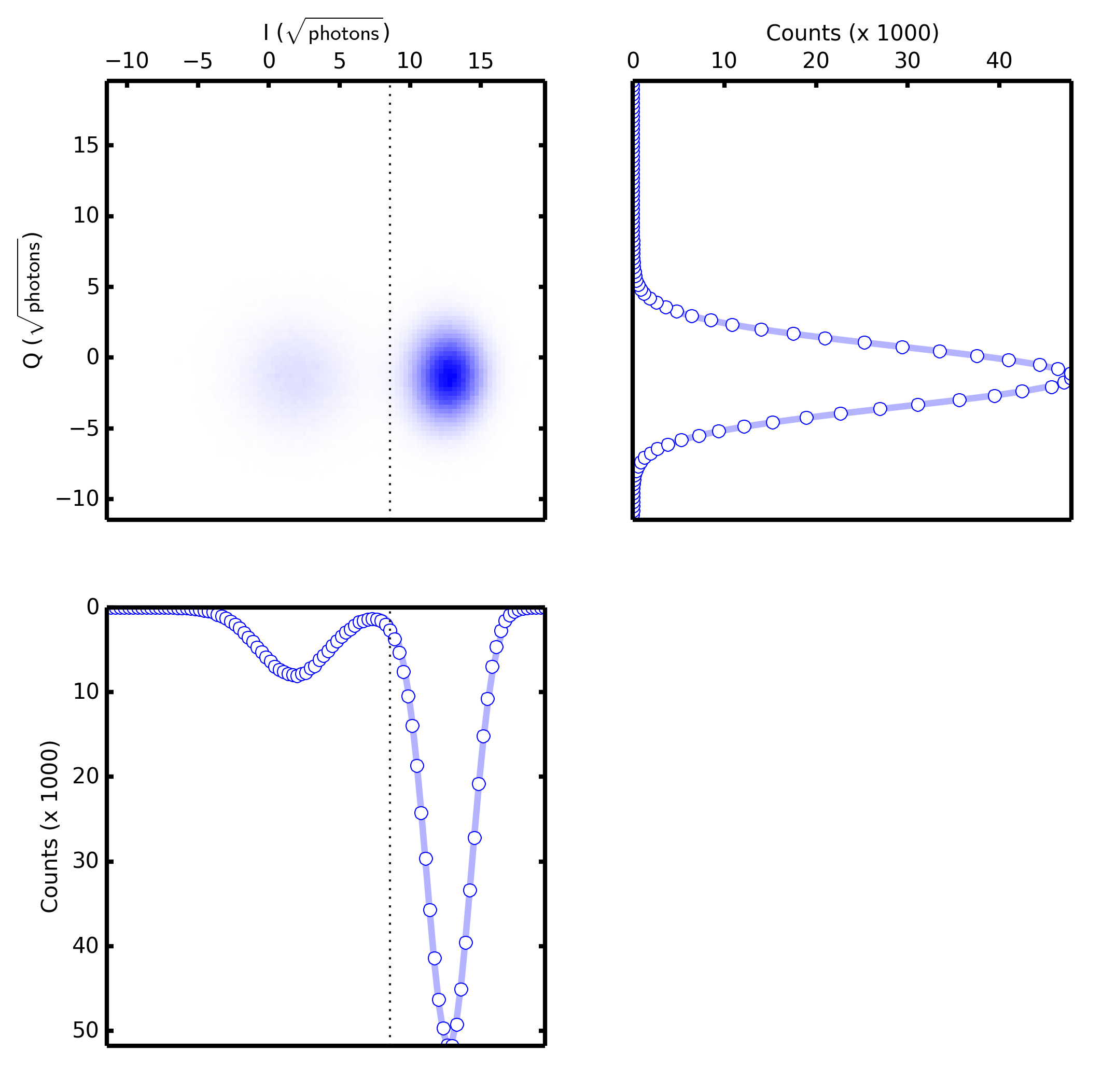}
\caption{Single shot readout of the qubit state with the JPC. Top left panel: two-dimensional histogram of the ($I$,$Q$) values of 820,000 measurements of the qubit in thermal equilibrium (20\% ground state and 80\% excited state). This histogram was rotated such that the information about the qubit state is encoded in the $I$ quadrature. The right and left gaussian distributions correspond to the qubit in $\ket{g}$ and $\ket{e}$ respectively. Bottom panel: histogram of the $I$ values, where the sum of two gaussians (full line) is fitted to the data (full dots). Right panel: Histogram of the $Q$ values, where a single gaussian (full line) is fitted to the data (full dots). The dotted line is the measurement threshold: if a data point lies on the left or right of this threshold, the outcome is associated with $\ket{e}$ or $\ket{g}$ respectively. The right gaussian is squeezed in the $I$ quadrature due to the amplifier saturation. }
\label{fig:JPC}
\end{figure}

\clearpage

\begin{figure}[ht!]
\setlength{\unitlength}{1cm}
\includegraphics[width=\columnwidth]{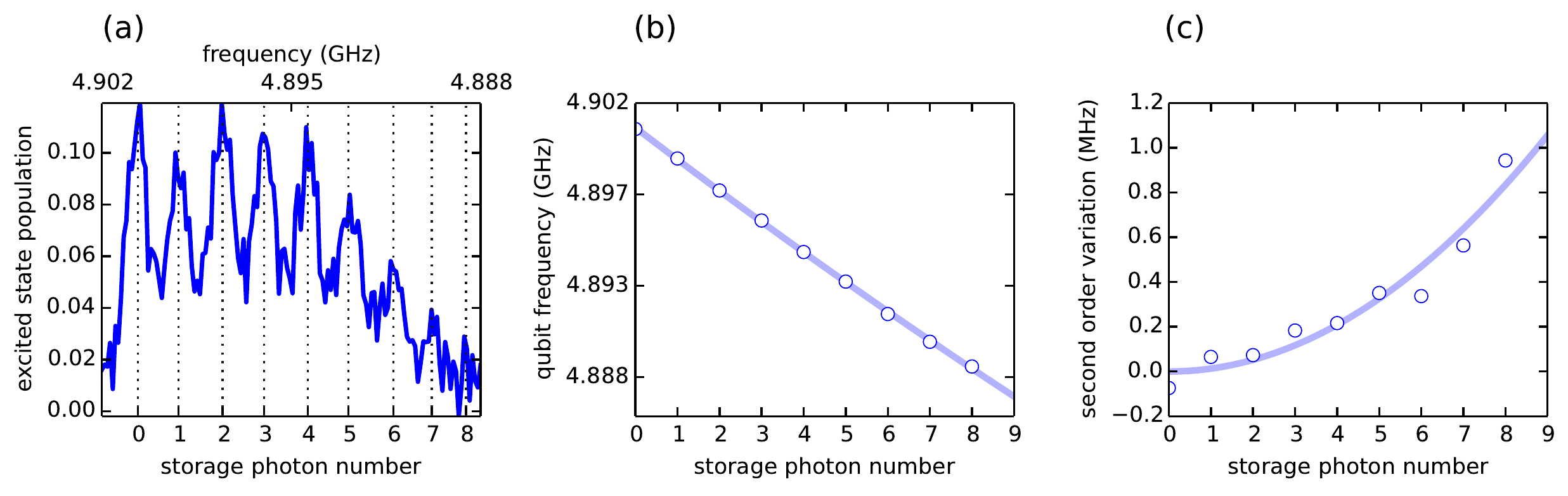}
\caption{Qubit frequency as a function of the number of photons in the storage cavity. When there are n=0 photons in the storage, the qubit frequency is $f_q=4.9007~$ GHz. As we introduce exactly $n$ photons in the storage, the qubit frequency shifts in discrete steps to $f_n=f_q-\frac{\chi_{rq}}{2\pi}n+\frac{\chi_{rq}^{(3)}}{2\pi}n^2$. We fit the data (full dots) to a quadratic function of n (full line), which gives the quantities $\chi_{qr}/2\pi=1.585~$MHz, and $\chi_{rq}^{(3)}/{2\pi}=5~$kHz \cite{Sun2014}. This qubit spectroscopy experiment was performed after the storage reached the statistical mixture of $\pm\alpha_\infty$ after 19 $\mu$s of pumping (see last panel of Fig.~\ref{fig:Numbersplitting_time_evolution}(a)). }
\label{fig:qubitnumbersplitting}
\end{figure}

\clearpage

\begin{figure}[ht!]
\setlength{\unitlength}{1cm}
\begin{picture}(18,12)
\put(0,6){\includegraphics[width=\columnwidth]{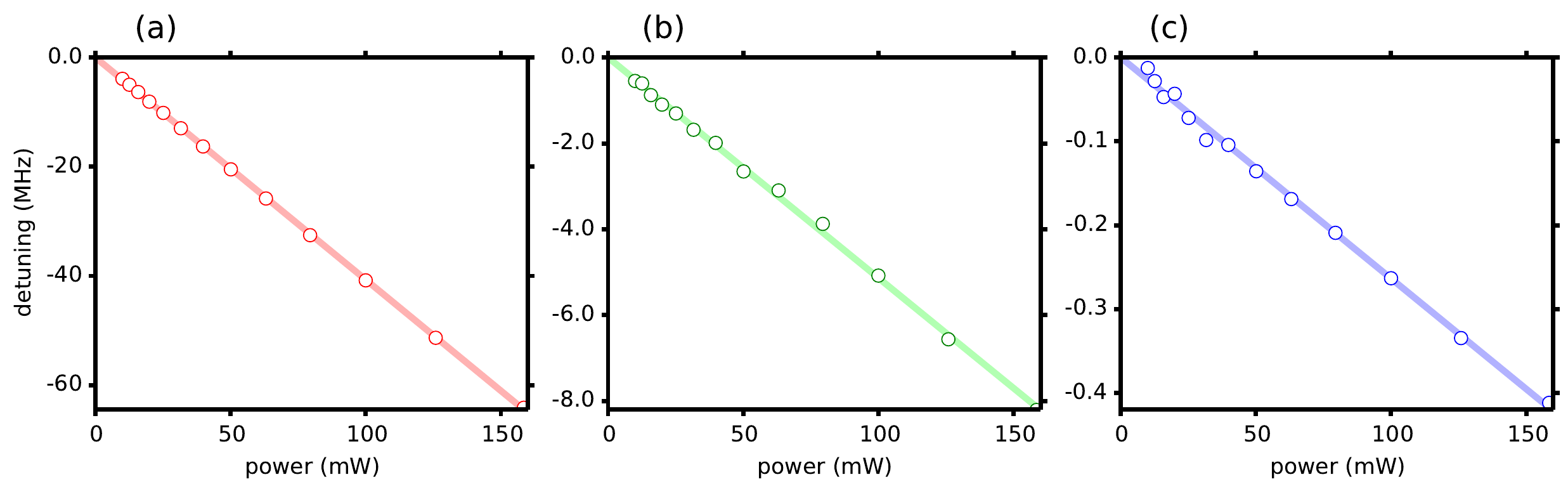}}
\put(0,0){\includegraphics[width=\columnwidth]{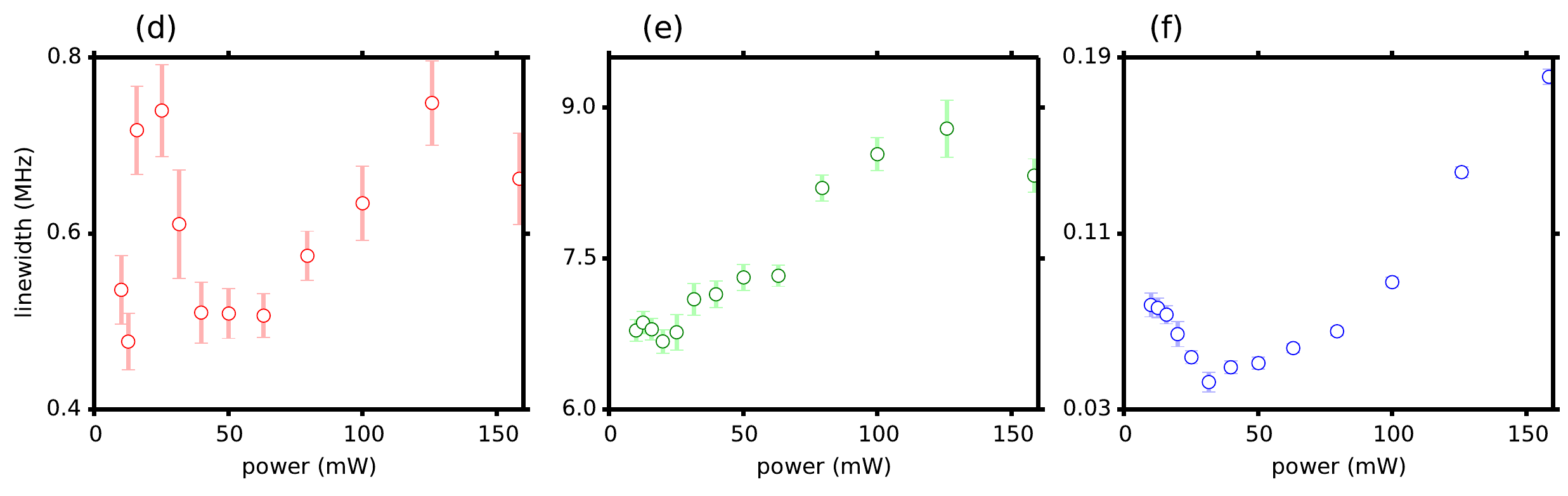}}
\end{picture}
\caption{AC stark shift due to the pump tone. We place the pump tone at $\omega_p=8.011~$GHz, and varie its power. For each power, we measure the spectrum of the qubit (a,d), readout mode (b,c) and storage mode (c,e). The frequencies of these modes (a-c) decrease linearly with the pump power, as shown by the linear fit (full line) to the data (full dots). The linewidths are represented in panels (d-f).}
\label{fig:Stark}
\end{figure}

\clearpage

\begin{figure}[ht!]
\setlength{\unitlength}{1cm}
\includegraphics[width=0.8\columnwidth]{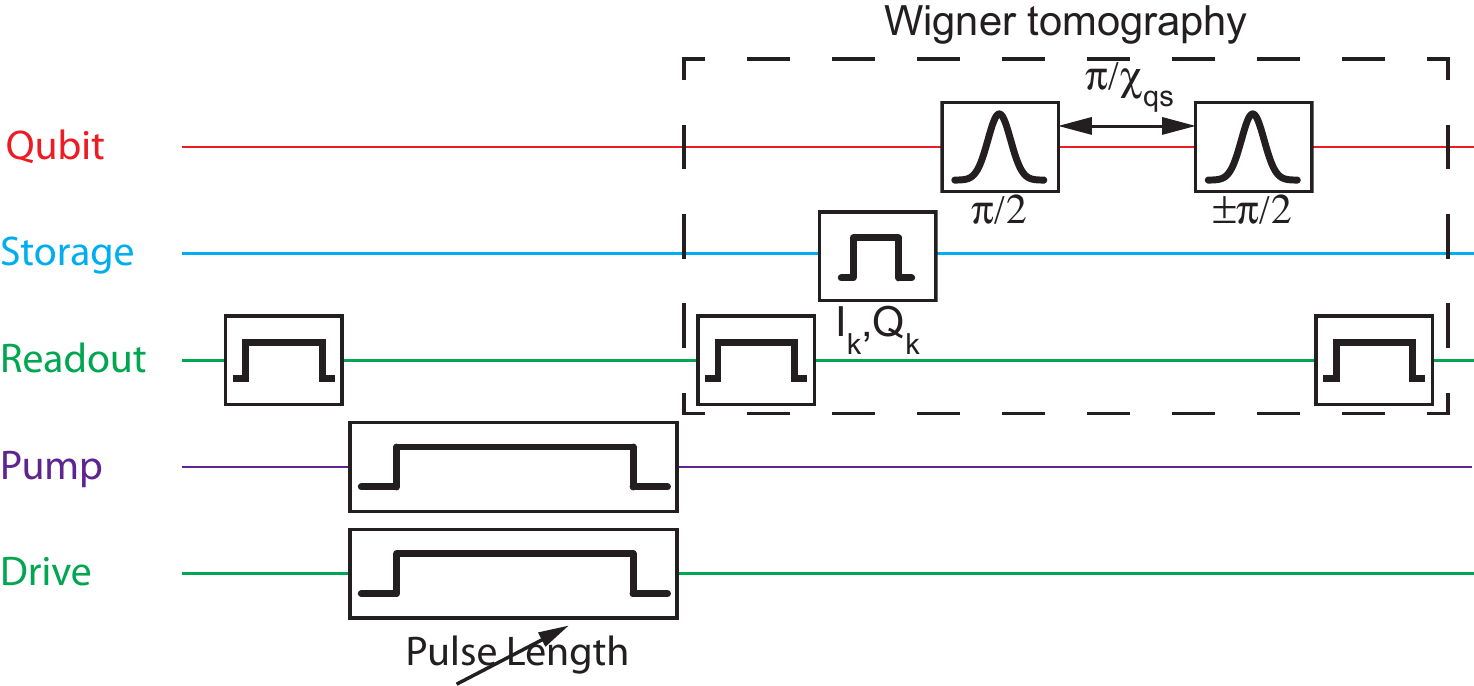}
\caption{Pulse sequence which generates the data of Fig.~\ref{fig:time_evolution}. First we initialize the qubit state by measuring it and post-select on it being in its ground state. Then we switch the pump and drive on for a variable amount of time. Finally, we perform Wigner tomography. The pulse sequence corresponding to the tomography is in the dashed rectangle and is described in Section~\ref{sec:WignerTomo}.}
\label{fig:pulseSeq}
\end{figure}

\clearpage

\begin{figure*}[ht!]
\setlength{\unitlength}{1cm}
\begin{picture}(19,16)
\put(0.5,10){\includegraphics[width=\columnwidth]{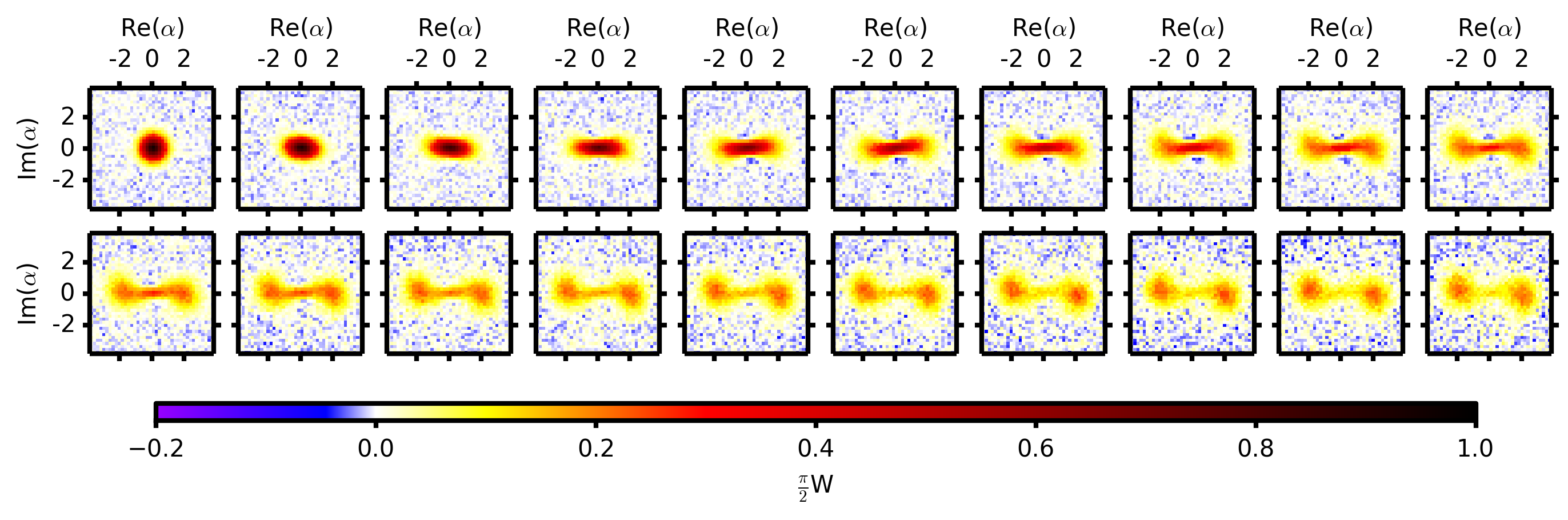}}
\put(0.5,5){\includegraphics[width=\columnwidth]{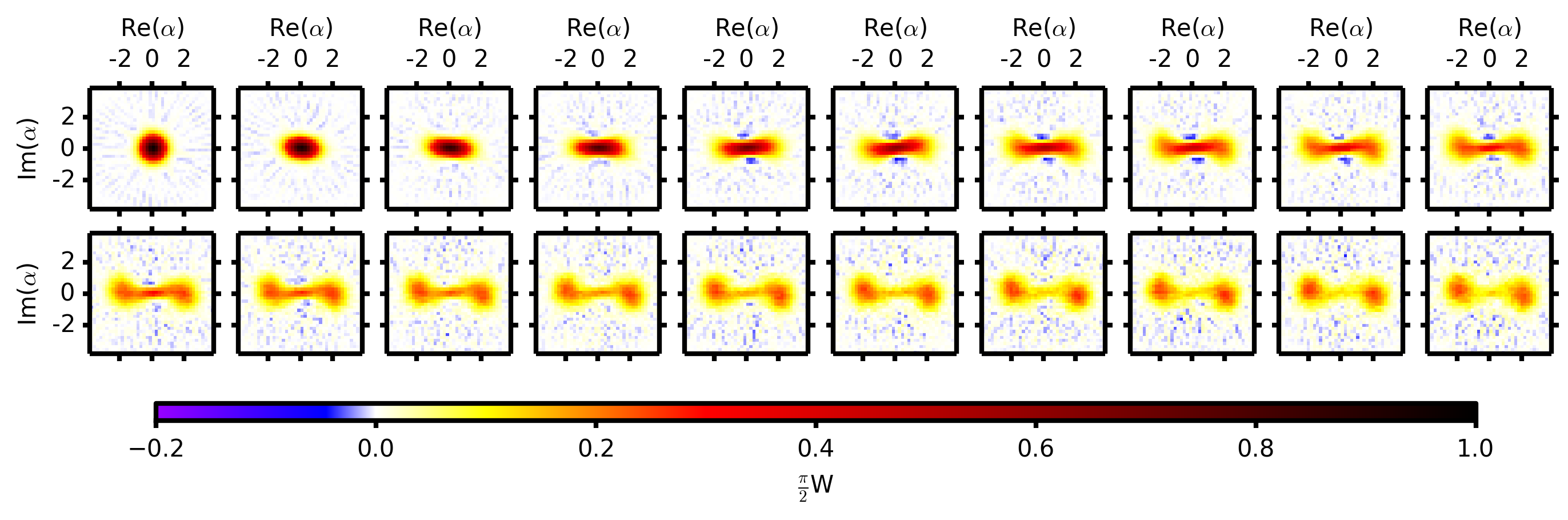}}
\put(0.5,0){\includegraphics[width=\columnwidth]{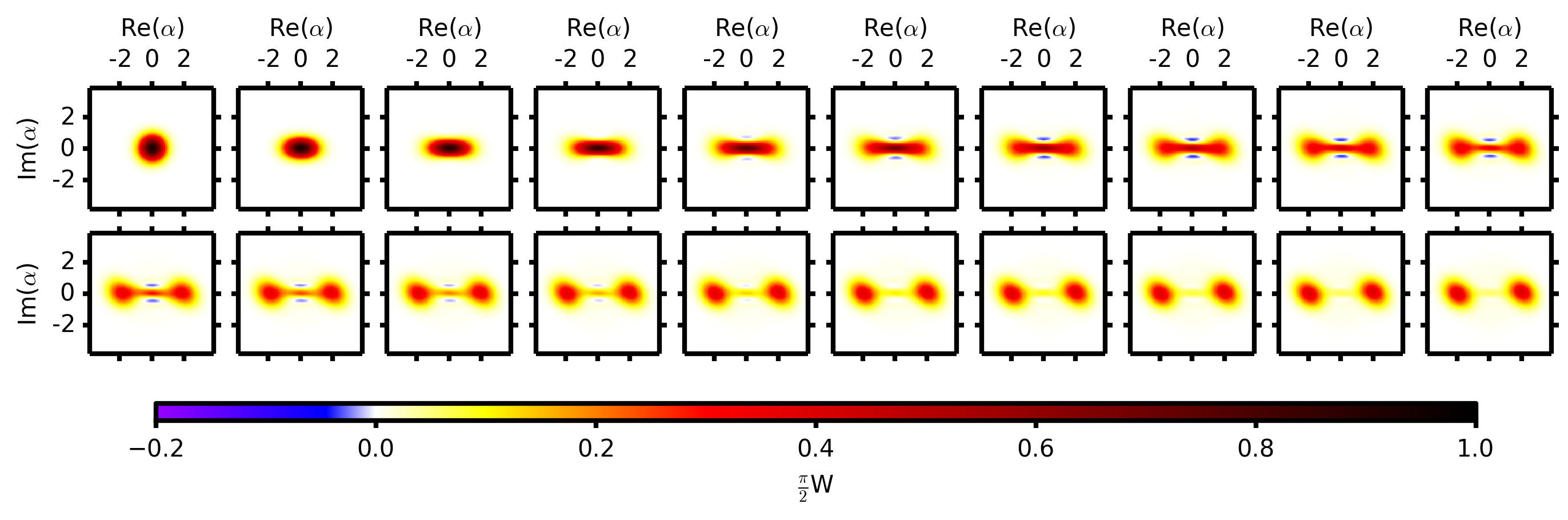}}
\put(0.5,15.5){(a)}
\put(0.5,10.5){(b)}
\put(0.5,5.5){(c)}
\end{picture}
\caption{Evolution of the storage mode state during pumping. We initialize the storage state in vacuum and switch on the pump and drive tones for various times $t_k$. Each one of the 20 panels in (a-d), ordered from left to right and top the bottom, is the Wigner function of the storage state after $t_k=k~\mu$s of pumping. We compare of the raw data (a) to the Wigner functions obtained from a reconstructed density matrix \cite{Vlastakis2013} (b) and from numerical simulations (c). The strong resemblance between the raw data and the Wigner from reconstructions demonstrates that our data can indeed be reproduced by a physically allowed density matrix: that is a positive, hermitian and trace one matrix. This shows that any systematic errors are reasonably low. The numerical simulations are obtained from solving the Lindblad master equation with Hamiltonian \eqref{eq:totalH1}\eqref{eq:totalH2}, assuming the drive and pump tones have well tuned frequencies, i.e. $\bH_\text{shift}=0$. We include photon loss and thermal processes for all three modes. All parameters included in this simulation were independently measured or estimated.}
\label{fig:time_evolution}
\end{figure*}

\clearpage

\begin{figure*}[ht!]
\setlength{\unitlength}{1cm}
\begin{picture}(18,8.1)
\put(0.5,4){\includegraphics[width=\columnwidth]{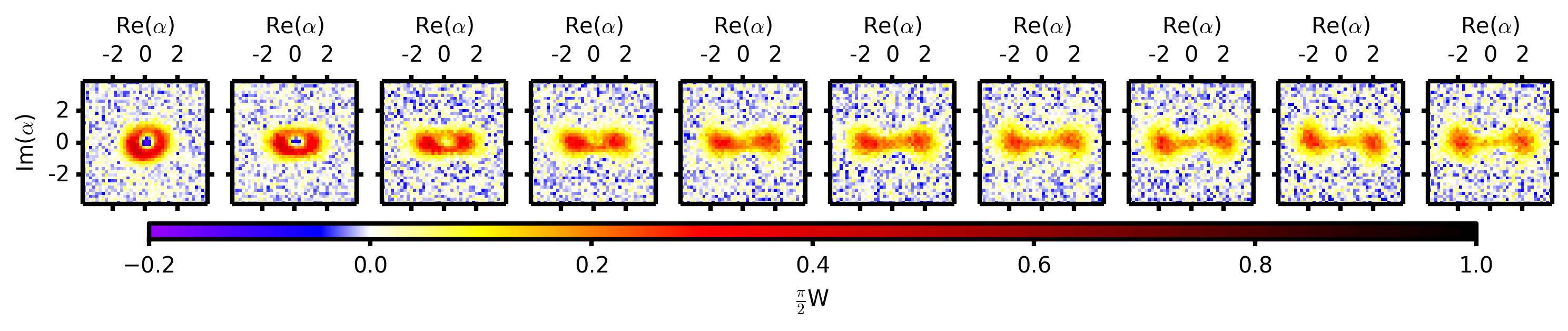}}
\put(0.5,0){\includegraphics[width=\columnwidth]{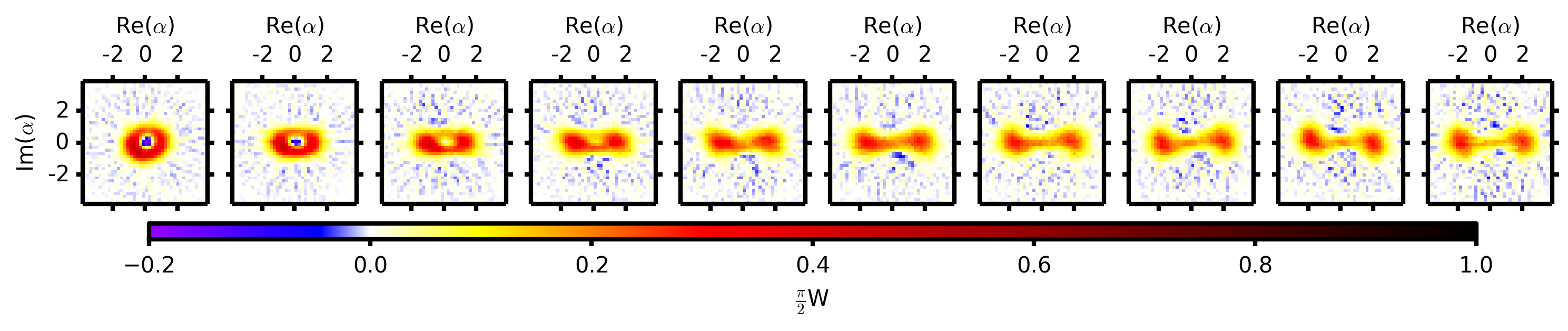}}
\put(0.5,3){(b)}
\put(0.5,7.5){(a)}
\end{picture}
\caption{Evolution of the storage mode state during pumping. We initialize the storage state in Fock state $\ket{1}$ and switch on the pump and drive tones for various times $t_k$. Each one of the 10 panels in (a,b), ordered from left to right, is the Wigner function of the storage state after $t_k=k~\mu$s of pumping. We compare of the raw data (a) to the Wigner functions obtained from a reconstructed density matrix (b). The Fock state is prepared by displacing the storage mode by a coherent state with an average photon number of 0.5, and then projecting to the odd parity manifold by measurement \cite{Sun2014}. As in Fig.~\ref{fig:time_evolution}, the state starts by squeezing in the $Q$ quadrature. At $t=3~\mu$s, the state resembles an odd Schr\"{o}dinger cat state where a cut of the Wigner function at $I=0$ alternates between 0, then positive, 0 at the center (this would be negative in the ideal lossless case), positive, and finally 0 again. Indeed, since we initialize the storage mode in an odd parity state, its evolution under exchanges of photon pairs conserves parity, and hence the transient superposition state has odd parity. As in Fig.~\ref{fig:time_evolution}, the state finally converges to a classical mixture of the two pointer states centered around $\ket{\pm\alpha_\infty}$.}
\label{fig:Fock_time_evolution}
\end{figure*}

\clearpage

\begin{figure}[ht!]
\setlength{\unitlength}{1cm}
{\includegraphics[width=10cm]{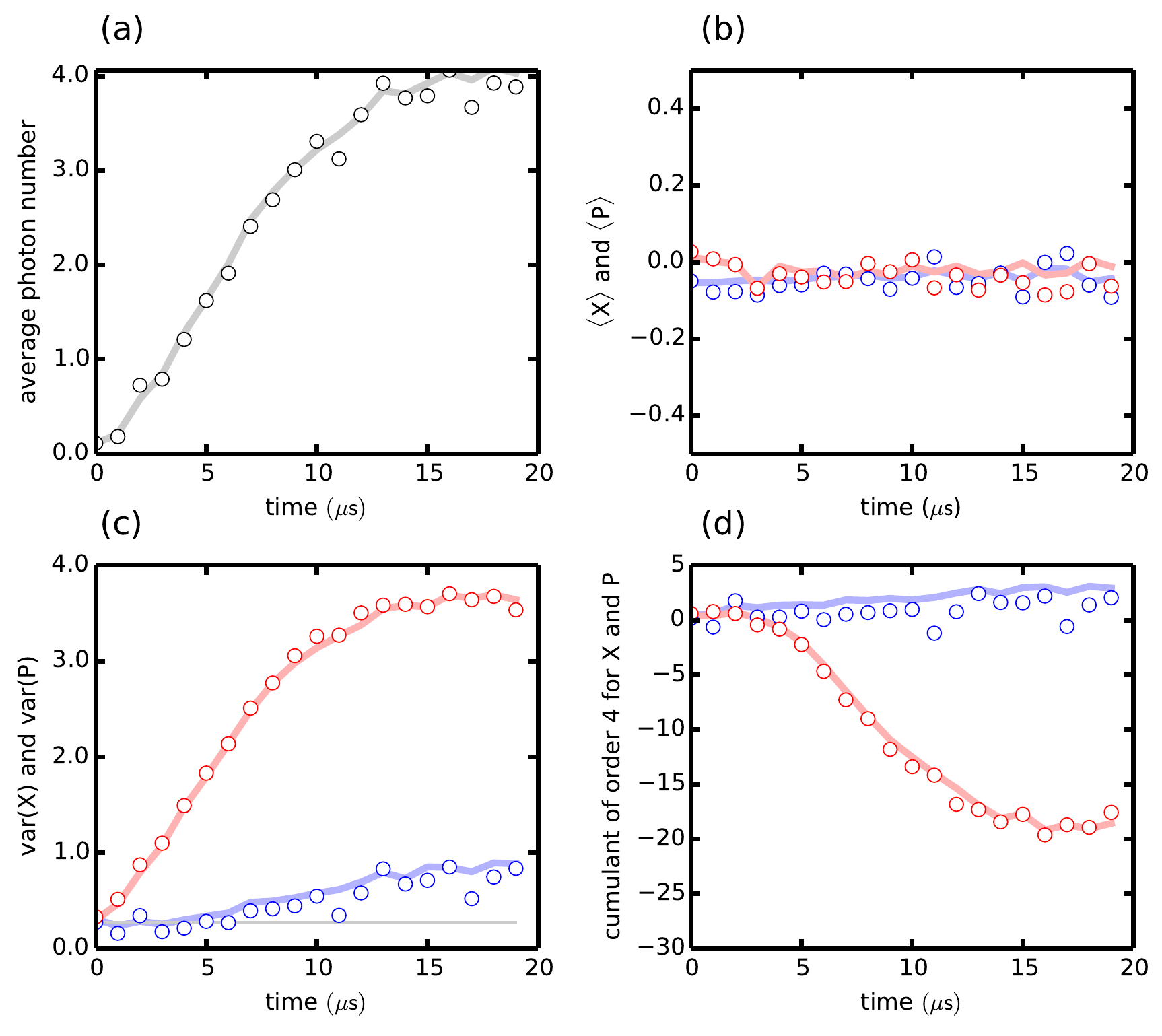}}
\caption{Expectation values of observables for the storage state during pumping. From the Wigner functions presented in Fig.~\ref{fig:time_evolution}, we can calculate the expectation value of any observable \cite{Cahill1969,Cahill1969a}. Values extracted from raw data are in full dots, and those extracted from the reconstructed Wigner functions are in full line. We represent the average photon number (a), the averages (b) and the variances (c) of $X=\frac{\ba_s+\ba_s^\dag}{2}, P=i\frac{\ba_s-\ba_s^\dag}{2}$. Defining $\bar X=X-\bket{X}$ and $\bar P=P-\bket{P}$, we represent in (d) the fourth order cumulants: $\bket{{\bar X}^4}-3\bket{{\bar X}^2}^2$ and $\bket{{\bar P}^4}-3\bket{{\bar P}^2}^2$.}
\label{fig:Obs_vs_time}
\end{figure}

\begin{figure}[ht!]
\setlength{\unitlength}{1cm}
{\includegraphics[width=10cm]{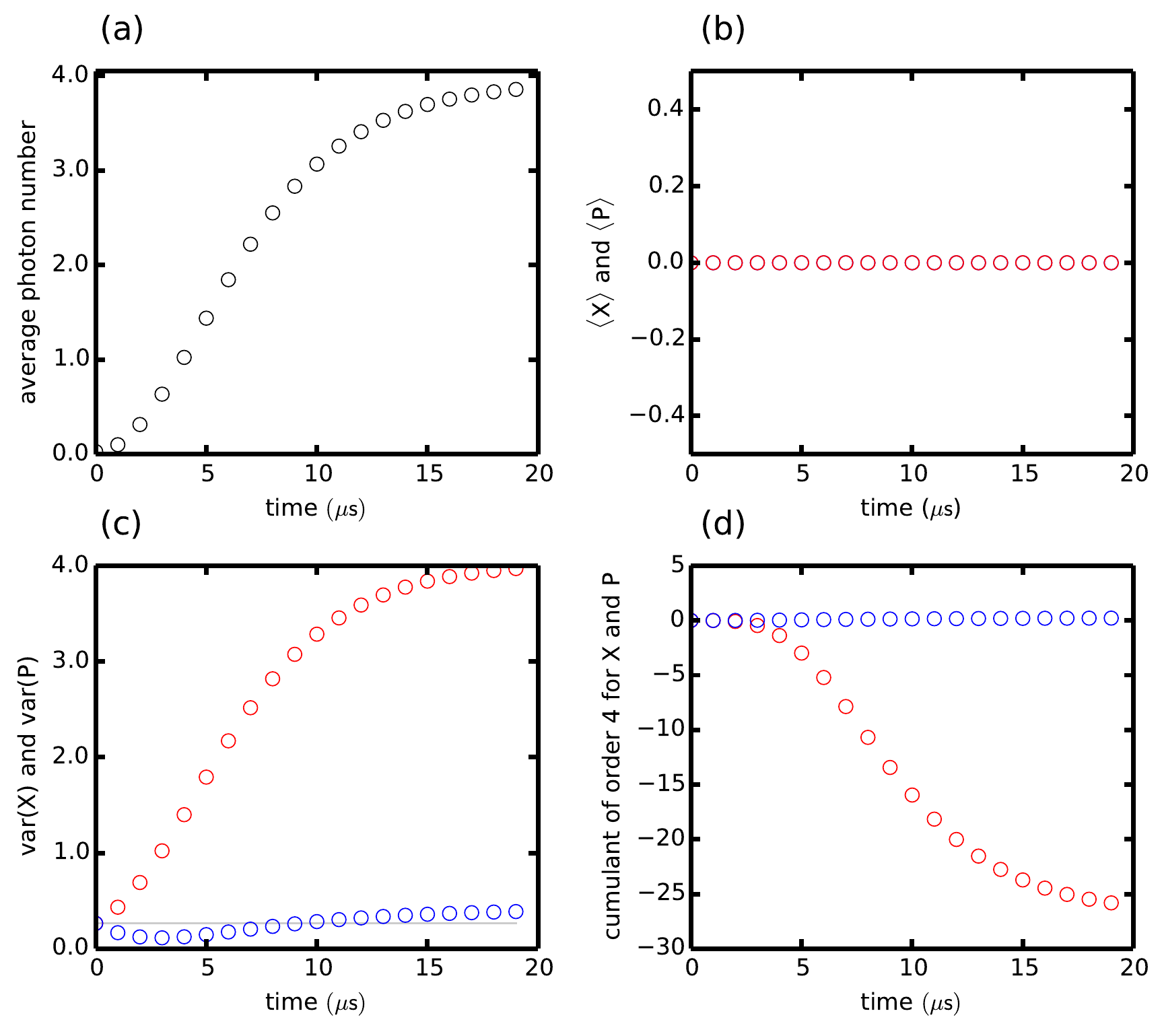}}
\caption{Identical description as Fig.~\ref{fig:Obs_vs_time}, where the values are extracted from the numerical simulations described in Fig.~\ref{fig:time_evolution}.}
\label{fig:Obs_vs_time_2}
\end{figure}

\clearpage

\begin{figure*}[ht!]
\setlength{\unitlength}{1cm}
\begin{picture}(18,12)
\put(0,6){\includegraphics[width=\columnwidth]{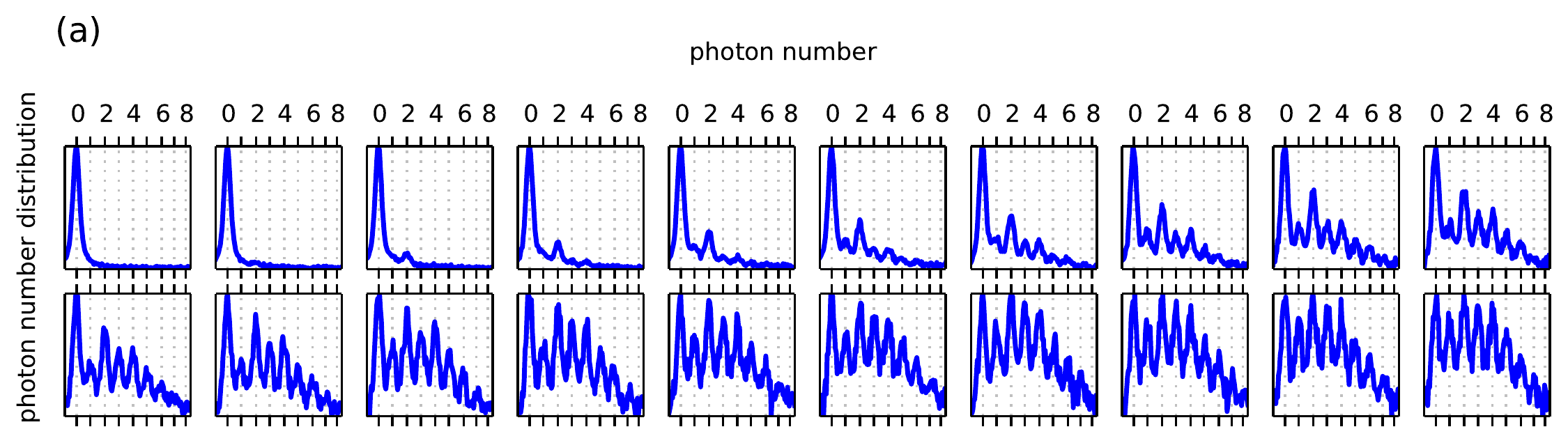}}
\put(0,0){\includegraphics[width=\columnwidth]{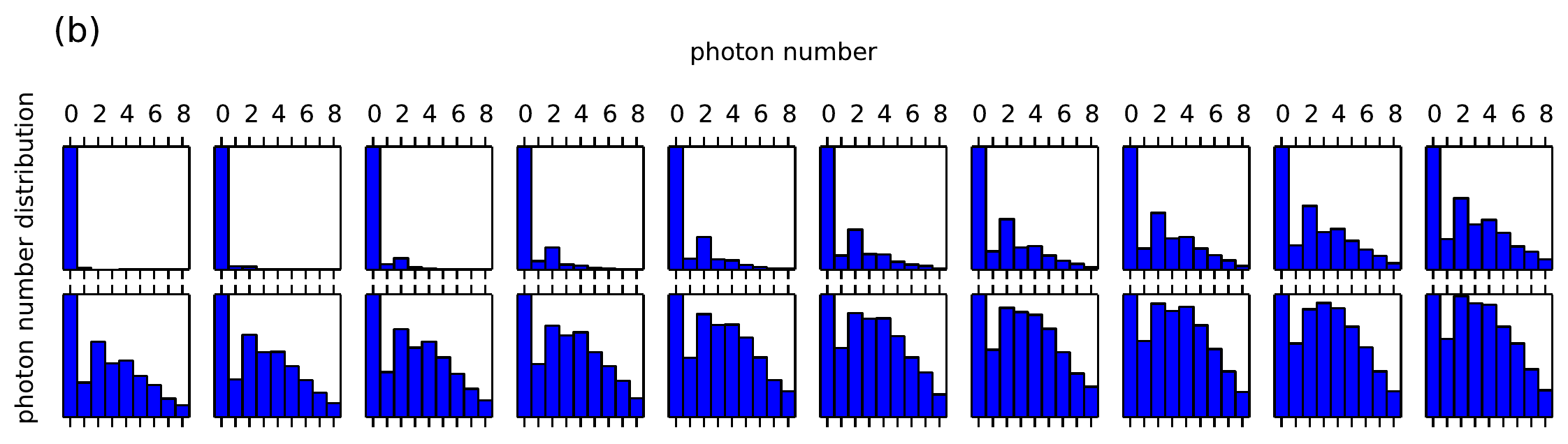}}
\end{picture}
\caption{Photon number distribution of the storage state during pumping. Each panel $k$ of the 20 panels in (a,b), ordered from left to right and top to bottom, represents the photon number distribution of the storage state after $t_k=k~\mu$s of pumping. In (a) we perform qubit spectroscopy with a $400~$ns sigma gaussian $\pi$ pulse. Due to the qubit-storage number splitting, this is a measure of the photon number distribution in the storage. In (b), we represent the diagonal of the reconstructed density matrix obtained from the Wigner tomography. These two independent measurements give consistent results, and exhibit the non-poissonian character of the photon number distribution during the transient evolution.}
\label{fig:Numbersplitting_time_evolution}
\end{figure*}

\end{document}